\documentclass[runningheads]{llncs}
\usepackage{graphicx}
\usepackage{amsmath}
\usepackage{amsfonts}
\usepackage{amssymb}
\usepackage{subfigure}
\usepackage{hyperref}
\usepackage{academicons}
\usepackage{xcolor}
\usepackage{booktabs}
\definecolor{rossopatavino}{RGB}{155,0,20}

\usepackage{etoolbox}

\usepackage{kantlipsum} 

\newcommand{\repthanks}[1]{\textsuperscript{\ref{#1}}}
\makeatletter
\patchcmd{\maketitle}
  {\def\thanks}
  {\let\repthanks\repthanksunskip\def\thanks}
  {}{}
\patchcmd{\@maketitle}
  {\def\thanks}
  {\let\repthanks\@gobble\def\thanks}
  {}{}
\newcommand\repthanksunskip[1]{\unskip{}}
\makeatother

\newcommand{\orcid}[1]{\href{https://orcid.org/#1}{\textcolor[HTML]{A6CE39}{\aiOrcid}}}

\begin{document}
\title{
Twitter Bots Influence on the Russo-Ukrainian War During the 2022 Italian General Elections}
\titlerunning{Twitter Bots Influence on Rus-Ukr War in 2022 Italian Elections}
%
\author{Francesco Luigi De Faveri\inst{1}\thanks{Authors contributed equally.\protect\label{X}}\orcidID{0009-0005-8968-9485} \and
Luca Cosuti\inst{1}\repthanks{X}\orcidID{0009-0000-9993-4695} \and
Pier Paolo Tricomi\inst{1,2}\thanks{Corresponding Author.}\orcidID{0000-0003-1600-835X}
\and
Mauro Conti\inst{1,2}\orcidID{0000-0002-3612-1934}
}
%
\authorrunning{De Faveri and Cosuti et al.}
%

\institute{University of Padua, Padua, Italy \\
\email{\{francescoluigi.defaveri,luca.cosuti\}@studenti.unipd.it\\ \{tricomi, conti\}@math.unipd.it} \and
Chisito S.r.l, Padua, Italy}

%
%
%
\maketitle              
\begin{abstract}
In February 2022, Russia launched a full-scale invasion of Ukraine. This event had global repercussions, especially on the political decisions of European countries. As expected, the role of Italy in the conflict became a major campaign issue for the Italian General Election held on 25 September 2022.
Politicians frequently use Twitter to communicate during political campaigns, but bots often interfere and attempt to manipulate elections. Hence, understanding whether bots influenced public opinion regarding the conflict and, therefore, the elections is essential.

In this work, we investigate how Italian politics responded to the
Russo-Ukrainian conflict on Twitter and whether bots manipulated public opinion before the 2022 general election. We first analyze 39,611 tweets of six major political Italian parties to understand how they discussed the war during the period February-December 2022. Then, we focus on the 360,823 comments under the last month's posts before the elections, discovering around 12\% of the commenters are bots. By examining their activities, it becomes clear they both distorted how war topics were treated and influenced real users during the last month before the elections.

\end{abstract}
\section{Introduction}
\label{sec:introduction}

At the dawn of 24 February 2022, the president of the Russian Federation, Vladimir V. Putin, announced an imminent ``Special Military Operation''\ in the oriental part of Ukraine. Soon thereafter, the global political leaders decided which side to support in the Russo-Ukrainian conflict. Along with most European countries, Italian politics sided with Ukraine by approving a law decree on 28 February 2022~\cite{DL_28_Feb_22}. 
The consequences of this decision were numerous. For instance, Italy reported a massive increase (+138\%) of cyber-attacks directed at critical infrastructures, apparently caused by hackers lined up with Russia~\cite{Report_Sole24h}. Additionally, Italian public opinion soon divided over the modalities of supporting Ukraine, such as sending military aid or applying sanctions to Russia. Since international relations inevitably impact democratic domestic politics~\cite{gaubatz1999elections}, the role of Italy in the Russo-Ukrainian conflict was a major campaign issue for the Italian (snap) general election on 25 September 2022.

\begin{figure}[hbtp]
        \centering
            \subfigure[Original tweet] 
            {
                \label{original_tweet}
                \includegraphics[width = 0.47\textwidth]{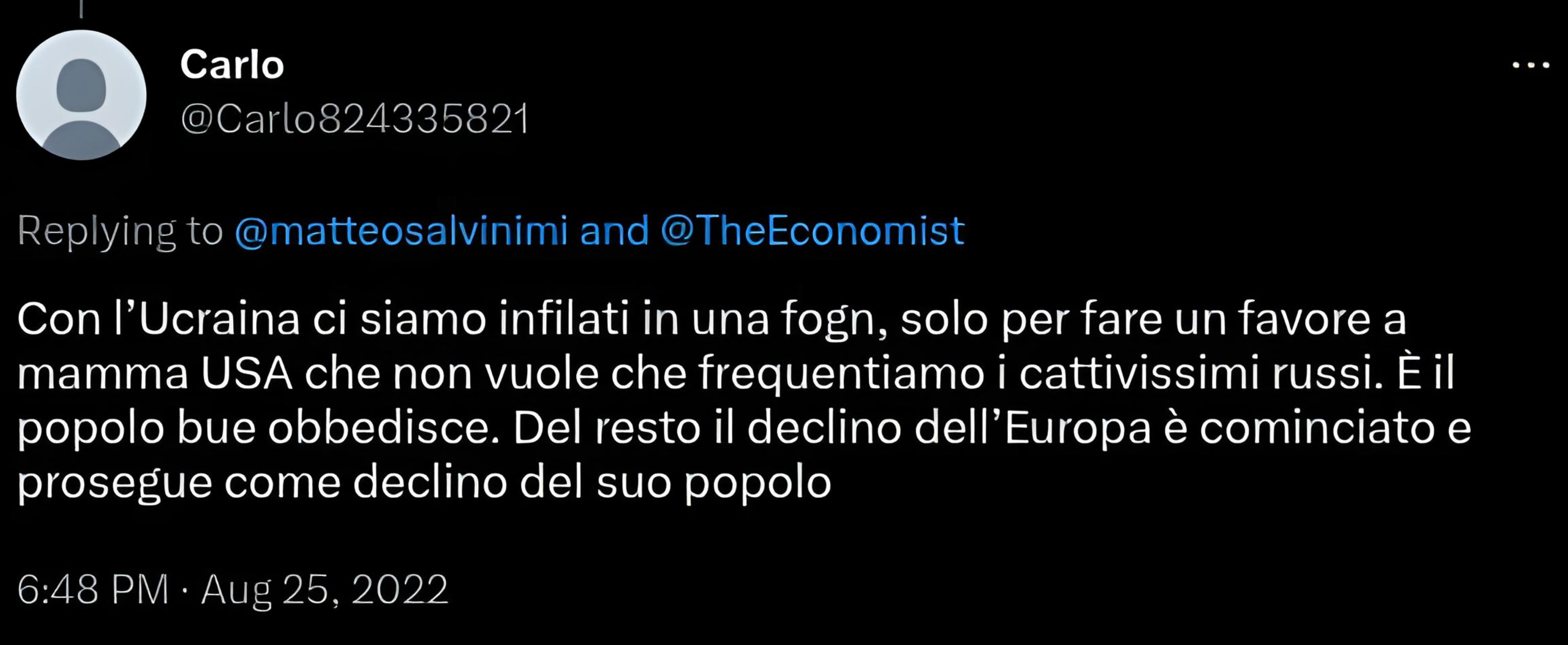} 
            } 
            \subfigure[Translated version] 
            {
                \label{transleted_tweet}
                \includegraphics[width = 0.47\textwidth]{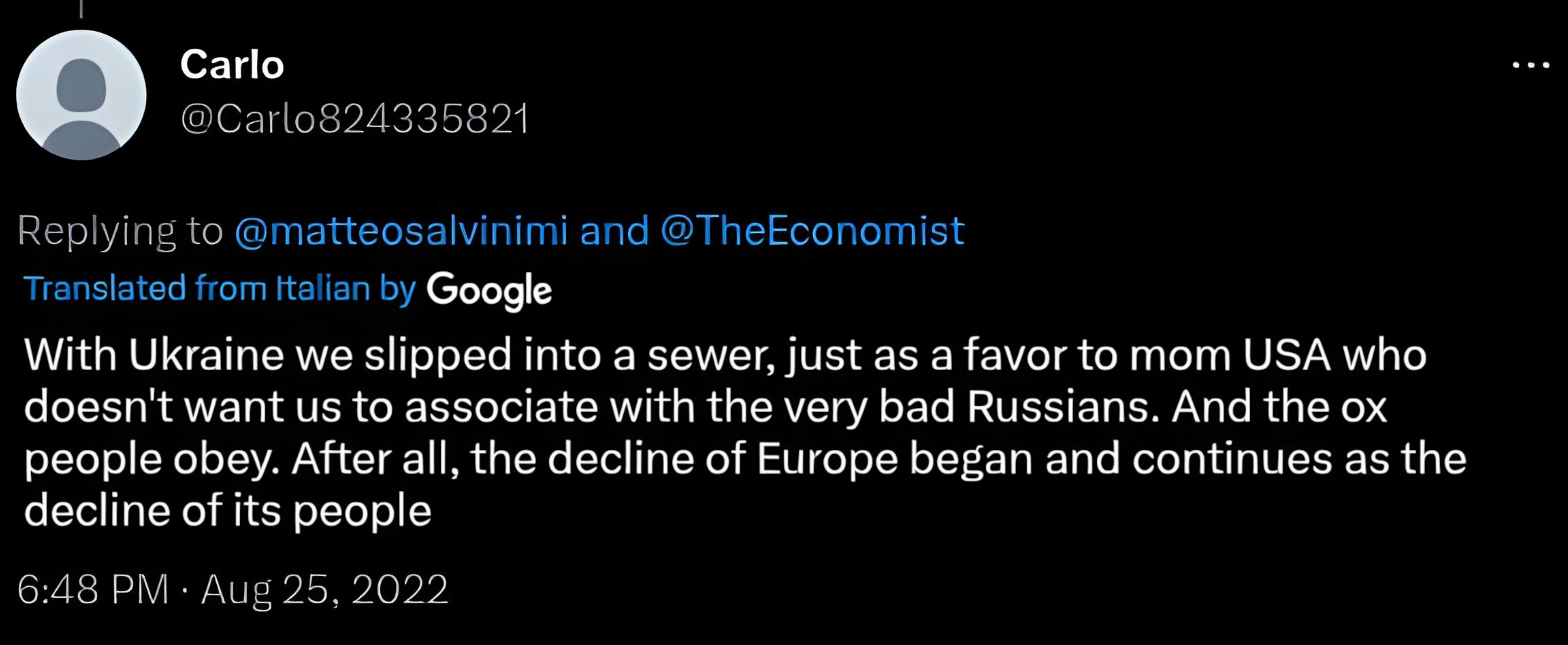} 
            }
         \caption{Bot response to an Italian politician expressing a strong-sided opinion regarding the conflict.}
        \label{fig:bot_tweet_example}
\end{figure}

People and politicians started expressing their concerns and opinions regarding the Russo-Ukrainian war on social media platforms like Facebook~\cite{caravaca2022estimating}, TikTok, Instagram, and Twitter. 
As largely demonstrated in the literature, opinions on social media are often manipulated by social bots~\cite{woolley2016automating,vasilkova2019social,bardi2023social} or colluding activities~\cite{dutta2019blackmarket,tricomi2022we}. 
Clear evidence has been found, for instance, in Japan's 2014 general election~\cite{schafer2017japan} or USA presidential elections in 2016~\cite{linvill2019russians} and 2020~\cite{SocialBots2020US}. Presumably, the last Italian general elections have not been exempted. Figure~\ref{fig:bot_tweet_example} illustrates a bot's provocative tweet in response to Matteo Salvini, a leader of Italian politics. Therefore, studying the impact of bots is fundamental for understanding the potential consequences they may have on social dynamics and online interactions. By investigating the role of bots in shaping the community, we can gain valuable insights into how they may have influenced the dissemination of information and the formation of opinions.

\paragraph{Contribution.} In this work, we investigate how Italian politics responded to the Russo-Ukrainian conflict on Twitter and whether bots manipulated public opinion before the 2022 general elections. In particular, we collected 39,611 tweets made by members of the main 6 political parties that belong to a left-wing or right-wing coalition from the period February-December 2022. We first conduct a semantic and temporal analysis of how politicians discussed the war, showing that some parties showed a high level of interest in the conflict and were actively engaged in commenting on the issue while others remained relatively silent. Secondly, we analyze 360,823 comments made during the last month of the political campaigns, from 23 August 2022 to 23 September 2022, examining bots' activities and influences on genuine users. We detected bots using Botometer~\cite{yang2022botometer}, a popular tool capable of evaluating the realness of an account using a Machine Learning-based classification method. Our results show that around 12\% of the profiles commenting on political posts are bots. Particularly, we found that bots have manipulated topics related to the Russo-Ukrainian war, especially on the center-right coalition, and that they influenced real users, often driving o soliciting discussions related to the conflict. 
We summarize our contributions as follows:
\begin{enumerate}
    \item[\textbullet] We collected a dataset of 39,611 tweets posted between 24 February 2022 and 31 December 2022, from the six major parties in Italy, and 360,823 comments from 105,603 unique users who replied during the last month of the 2022 Italian general elections. The dataset will be made publicly available for future research;
    \item[\textbullet] We provide a detailed analysis of how the 6 major Italian parties expressed and sided concerning the Russo-Ukrainian war on Twitter from the beginning of the war to the end of 2022;
    \item[\textbullet] We examine the bots' impact on Twitter and how they influenced real users regarding the Russo-Ukrainian war during the last month of the general elections.
\end{enumerate}

\paragraph{Organization.}
Section~\ref{sec:related works} discusses related works, while Section~\ref{sec:dataset} presents the dataset used in the experiments. In Section~\ref{sec:analysis} and Section~\ref{sec:bots}, we analyze politics in Italy during the conflict and the bots' influence on the elections, respectively. Section~\ref{sec:discussion} makes further discussion and Section~\ref{sec:concl} concludes the paper.
\section{Related Works}
\label{sec:related works}

In this section, we focus on the state-of-the-art analysis of bot infiltration in delicate scenarios and opinion manipulation through Twitter. Antonakaki et al.~\cite{antonakaki2021survey_twitter} conducted a comprehensive literature review presenting different approaches and techniques used for Twitter research. The authors acknowledged that Twitter had become a valuable data source for researchers, offering data for many purposes, such as forecasting social, economic, or commercial indicators~\cite{twit_forecast} as well as assessing and predicting political polarization~\cite{khan2021election,giakatos2023pypoll}.
For instance, Weber et al.~\cite{islam_polarization}, during the 2013 ``Arab Spring'' in Egypt, collected and analyzed a large dataset of tweets to categorize the users based on their political affiliation. 

However, such information is often undermined by the presence of bots, i.e., automated accounts used to engage and behave mimicking human users, often controlled by a bot master. While there are some benevolent social media bots, many are used for dishonest and nefarious purposes~\cite{aiello2012people,varol2022should}.
The existence of bots on the Twitter platform has been firmly established through many academic investigations~\cite{alothali2018detecting,chavoshi2016identifying,gilani2017bots,mannocci2022mulbot,cresci2020decade}, and news articles~\cite{Bots_percentage,Bots_early_stages}. 
Weng et al.~\cite{publicopinionmanipulation} explained the differences between the opinion manipulations done by bots compared with those from real users, and Mazza et al.~\cite{mazza2022investigating} investigated the difference between trolls, social bots, and humans on Twitter.
Notably, these accounts can wield an exceptionally strong influence in delicate situations~\cite{BotsInRussia}, such as stock trading \cite{cresci2019cashtag}, sensitive content diffusion~\cite{singh2016behavioral}, vaccination~\cite{vaccine_bots}, or political elections manipulation. 
Regarding the latest, Pastor et al.~\cite{pastor2020spotting} analyzed the presence and behavior of social bots on Twitter in the context of the November 2019 Spanish general election. They limited the analysis of the bots' interaction up to seven days before Election day using Social Feed Manager~\cite{SFM} to capture the tweets and analyze the bot. Fernquist et al.~\cite{fernquist2018political} presented a study on the influence of bots in the Swedish general election held in September 2018. Bessi and Ferrara~\cite{bessi2016social} investigated how the presence of social media bots impacted the 2016 Presidential elections in America, and similar works were conducted on the latest one in 2020~\cite{SocialBots2020US,ferrara2020characterizing}. For a comprehensive overview of bots, political elections, and social media, we refer to~\cite{ferrara2020bots}. 

\section{Dataset Creation}
\label{sec:dataset}

In this study, we collected our own Twitter dataset due to the unique nature of the analysis. We selected six parties to analyze according to the current political scenario in Italy. In particular, we considered:
\begin{enumerate}
    \item[\textbullet] The coalition that preceded Mario Draghi's technical government (the so-called ``giallo-rosso" government, who guided Italy from 5 September 2019 until 13 February 2021~\cite{secondo_governo_conte,conte2resignation}), made by the Democratic Party (Partito Democratico, PD), the Five Stars Movement (Movimento 5 Stelle, M5S)
    \item[\textbullet] The Italian Green-Left party (Sinistra Italiana-Verdi, SiVe);
    \item[\textbullet] The coalition that won the September 2022 elections, and is currently in power: Brothers of Italy (Fratelli d'Italia, FdI), League for Salvini Premier (Lega per Salvini Premier, Lega), and Forward Italy (Forza Italia, FI).
\end{enumerate} 
We then model each of the parties to be constructed as:
\begin{equation*}
    D_i = \left[P, L, p_1, \dots, p_6\right]
\end{equation*}
where:
\begin{enumerate}
    \item[\textbullet] $D_i$ is the Dataset, $i=1,\dots, 6$, one for each party.
    \item[\textbullet] $P$ is the \lq\lq Party account\rq\rq, e.g.,  @FratellidItalia.
    \item[\textbullet] $L$ is the \lq\lq Leader account\rq\rq, e.g.,  @GiorgiaMeloni.
    \item[\textbullet] $p_{1},\dots,p_{6}$ are six ``major political figures'' in that party, e.g., @Ignazio\_LaRussa, @DSantanche, @FrancescoLollo1, @FidanzaCarlo, @fabiorampelli and @isabellarauti.
\end{enumerate}
The final dataset has been constructed by collecting all the tweets from the party account, the leader account, and six other politicians in the party (following the structure defined above) that were posted from 24 February 2022 until 31 December 2022. To download the tweets, we queried the official Twitter API~\cite{TwitterAPI} to browse each profile's timeline and retrieve all the necessary tweets. 
After this initial collection of tweets, we focused on the posts published during the latest month of the political campaign in Italy, from 23 August 2022 until 23 September 2022. We considered all the content shared by the secretary of each party and every reply. 
An overview of the full dataset can be seen in Table~\ref{tab:Tab_3_full_dataset}. We indicate the party, the party leader, the selected profiles we fetched the information from, the cumulative number of followers of each party's profiles, and the overall number of posted tweets. For the last month of the political campaign, we considered all the content shared by the secretary of each party and every reply, as well as the number of unique commenters.
These numbers represent only the tweets directly posted by the party members. During the collection, we excluded the retweets to reduce the number of repeated tweets between different accounts, to avoid redundancy, and to have a real and clear opinion from each profile.

\begin{table}[htbp]
\centering
\caption{Complete overview of the dataset.}
\tiny
\begin{tabular}{llp{35mm}cccc}
\toprule
\textbf{\textit{Party}} & \textbf{\textit{Leader}} & \textbf{\textit{Members}} & \textbf{\textit{Total Followers}} & \textbf{\textit{Posted Tweets}} & \textbf{\textit{Replies to Secretary}} & \textbf{\textit{Unique Users Replying}} \\
\midrule
\href{https://twitter.com/pdnetwork}{PD} & \href{https://twitter.com/EnricoLetta}{Letta} & \href{https://twitter.com/serracchiani}{Serracchiani},  \href{https://twitter.com/AndreaOrlandosp}{Orlando}, \href{https://twitter.com/mariannamadia}{Madia}, \href{https://twitter.com/peppeprovenzano}{Provenzano}, \href{https://twitter.com/lauraboldrini}{Boldrini},\href{https://twitter.com/paologentiloni}{Gentiloni}. & $3.511M$ & $4357$ & $158747$ & $35571$ \\
\href{https://twitter.com/fratelliditalia}{FdI}   & \href{https://twitter.com/giorgiameloni}{Meloni} & \href{https://twitter.com/ignazio_larussa}{La Russa}, \href{https://twitter.com/dsantanche}{Santanchè}, \href{https://twitter.com/francescolollo1}{Lollobrigida}, \href{https://twitter.com/fidanzacarlo}{Fidanza}, \href{https://twitter.com/fabiorampelli}{Rampelli}, \href{https://twitter.com/isabellarauti}{Rauti}.  & $2.471M$              & $6610$          & $60237$                    & $22670$                 \\
\href{https://twitter.com/mov5stelle}{M5S}   & \href{https://twitter.com/giuseppeconteit}{Conte} & \href{https://twitter.com/roberto_fico}{Fico}, \href{https://twitter.com/paolatavernam5s}{Taverna}, \href{https://twitter.com/c_appendino}{Appendino}, \href{https://twitter.com/carlosibilia}{Sibilia}, \href{https://twitter.com/giuliagrillom5s}{Grillo}, \href{https://twitter.com/maiorinom5s}{Maiorino}.          & $2.419M$              & $3672$          & $47886$                    & $14255$                 \\
\href{https://twitter.com/legasalvini}{Lega}  & \href{https://twitter.com/matteosalvinimi}{Salvini} & \href{https://twitter.com/fontana3lorenzo}{Fontana}, \href{https://twitter.com/arrigoni_paolo}{Arrigoni}, \href{https://twitter.com/simopillon}{Pillon}, \href{https://twitter.com/edorixi}{Rixi},  \href{https://twitter.com/giamma71}{Centinaio}, \href{https://twitter.com/gbongiorno66}{Bongiorno}.       & $1.898M$              & $15797$         & $59317$                    & $20159$                 \\
\href{https://twitter.com/forza_italia}{FI}    & \href{https://twitter.com/berlusconi}{Berlusconi} & \href{https://twitter.com/antonio_tajani}{Tajani}, \href{https://twitter.com/berniniam}{Bernini}, \href{https://twitter.com/gasparripdl}{Gasparri}, \href{https://twitter.com/raffaelefitto}{Fitto}, \href{https://twitter.com/min_casellati}{Casellati}, \href{https://twitter.com/liciaronzulli}{Ronzulli}.    & $804.2K$               & $4172$          & $29597$                    & $9962$                  \\
\href{https://twitter.com/si_sinistra}{SiVe}  & \href{https://twitter.com/nfratoianni}{Fratoianni} & \href{https://twitter.com/angelobonelli1}{Bonelli}, \href{https://twitter.com/aboubakar_soum}{Soumahoro}, \href{https://twitter.com/chicoalemanni}{Alemanni}, \href{https://twitter.com/eleonoraevi}{Evi}, \href{https://twitter.com/giuliomarcon1}{Marcon}, \href{https://twitter.com/serpellegrino11}{Pellegrino}.    & $411 K$               & $5003$          & $5038$                     & $2986$                  \\
\bottomrule
\end{tabular}
\label{tab:Tab_3_full_dataset}
\end{table}

\section{The Russo-Ukrainian War in Italian Politics}
\label{sec:analysis}
We start our analysis by understanding whether and how frequently the Italian parties mentioned the Russo-Ukranian conflict (Section~\ref{sec:How did the politicians talk about the War?}). After that, we conduct a temporal analysis to determine when the conflict was primarily discussed, with a particular focus on election time (Section~\ref{sec:temporal-rus}).


\subsection{The Importance of Conflict for Italian Political Parties}
\label{sec:How did the politicians talk about the War?}
\begin{figure}[hbtp]
        \centering
            \subfigure[PD] 
            {
                \label{PD_wc}
                \includegraphics[width=.31\textwidth]{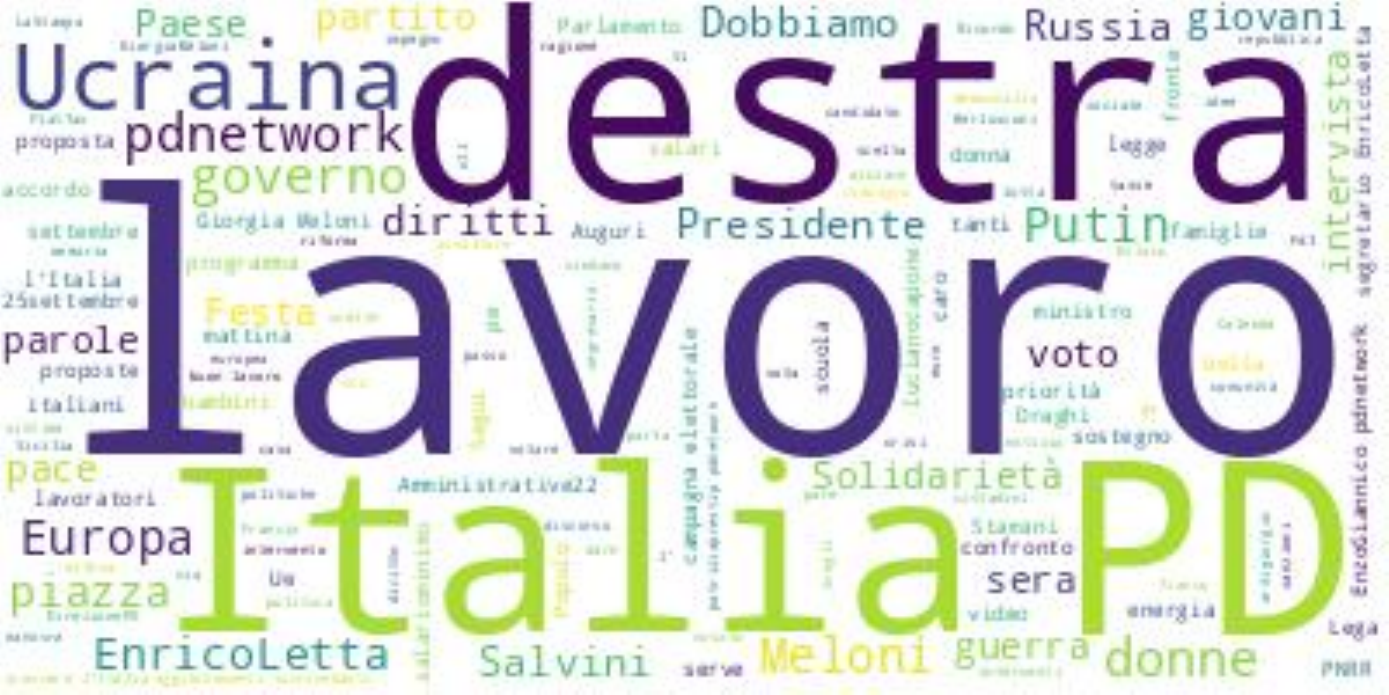} 
            } 
            \subfigure[M5S] 
            {
                \label{M5S_wc}
                \includegraphics[width=.31\textwidth]{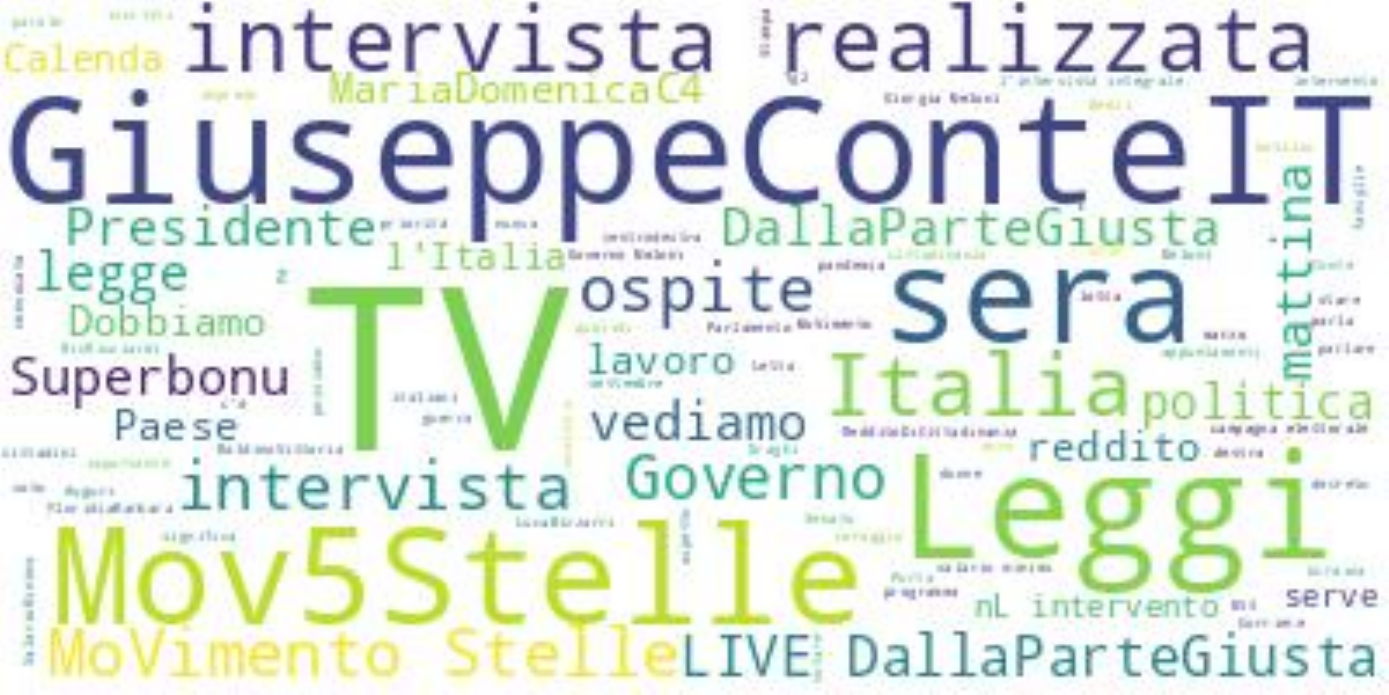} 
            }
            \subfigure[SiVe] 
            {
                \label{SiVe_wc}
                \includegraphics[width=.31\textwidth]{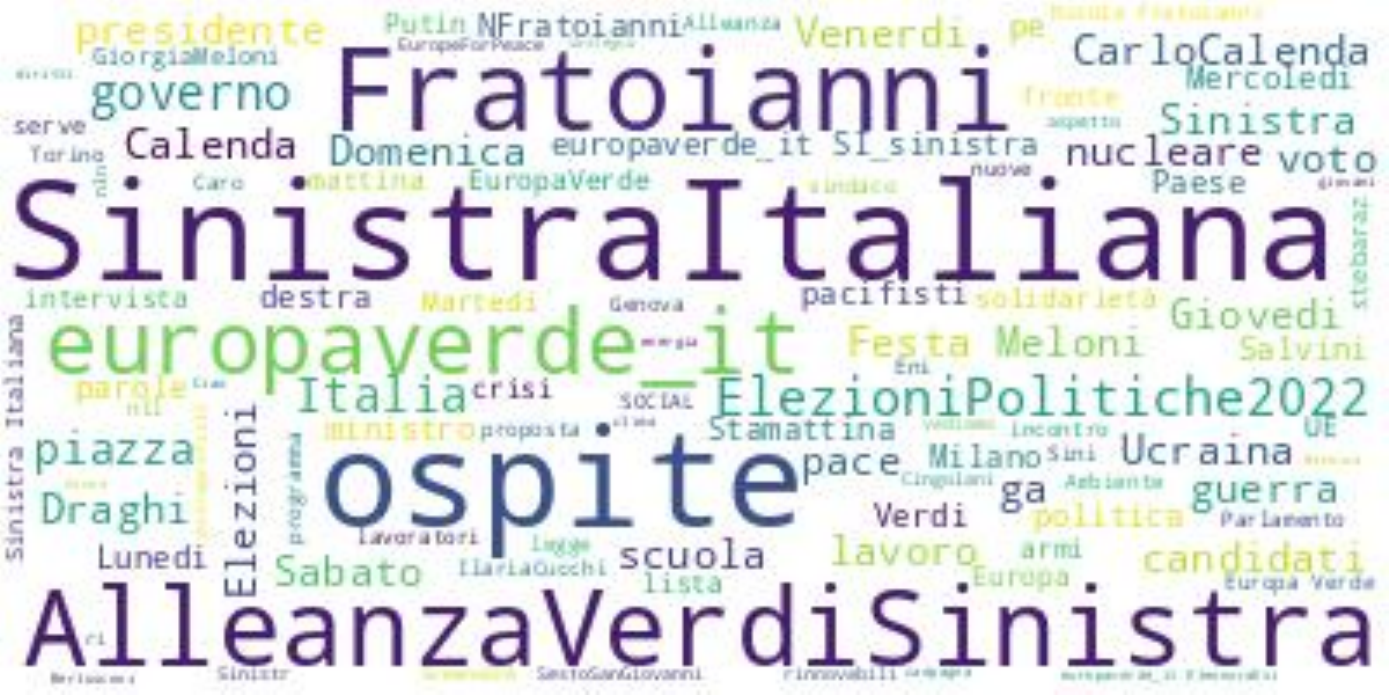} 
            }\\ 
            \subfigure[FdI] 
            {
                \label{FdI_wc}
                \includegraphics[width=.31\textwidth]{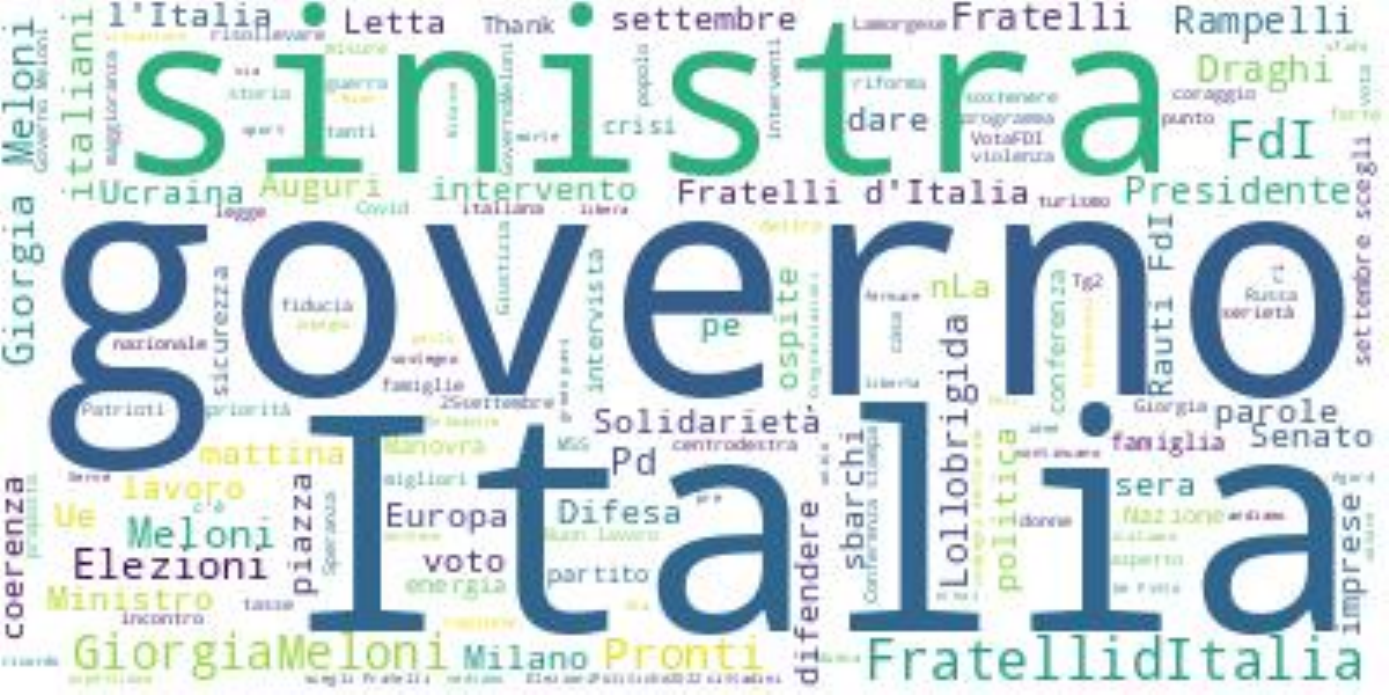} 
            }
            \subfigure[Lega] 
            {
                \label{Lega_wc}
                \includegraphics[width=.31\textwidth]{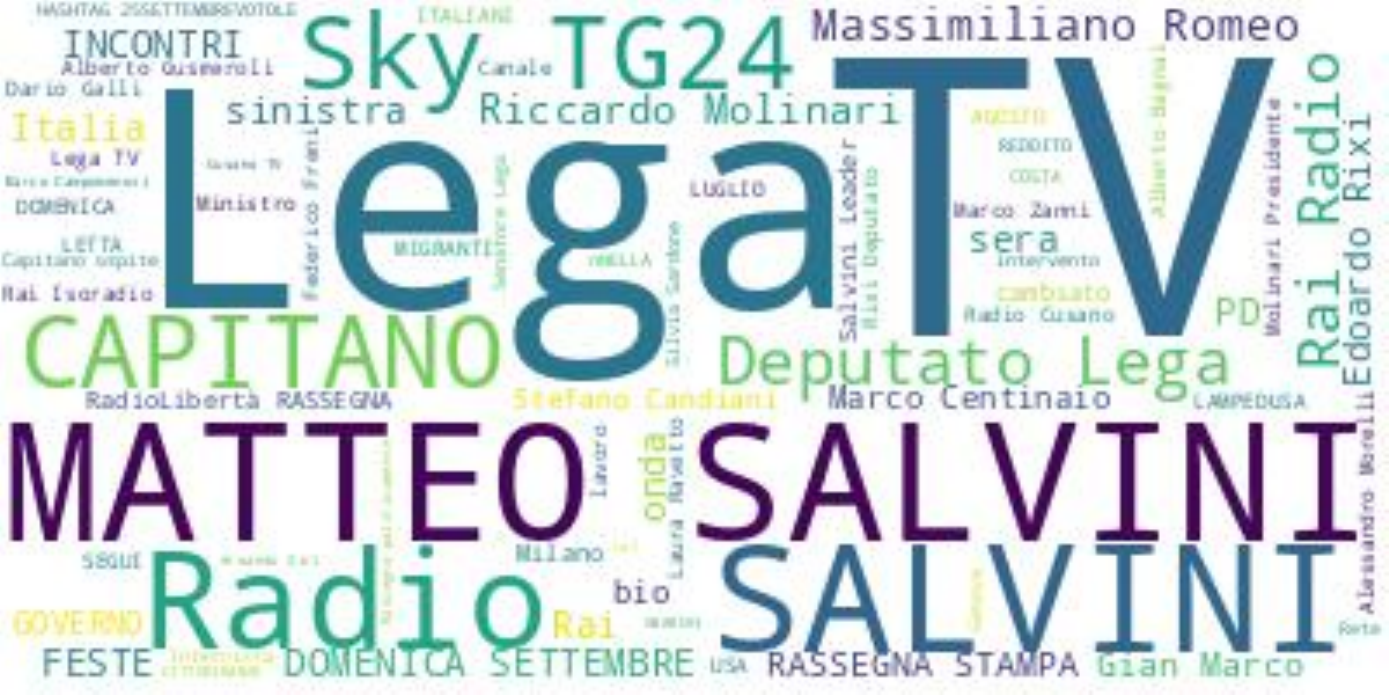} 
            }
            \subfigure[FI] 
            {
                \label{FI_wc}
                \includegraphics[width=.31\textwidth]{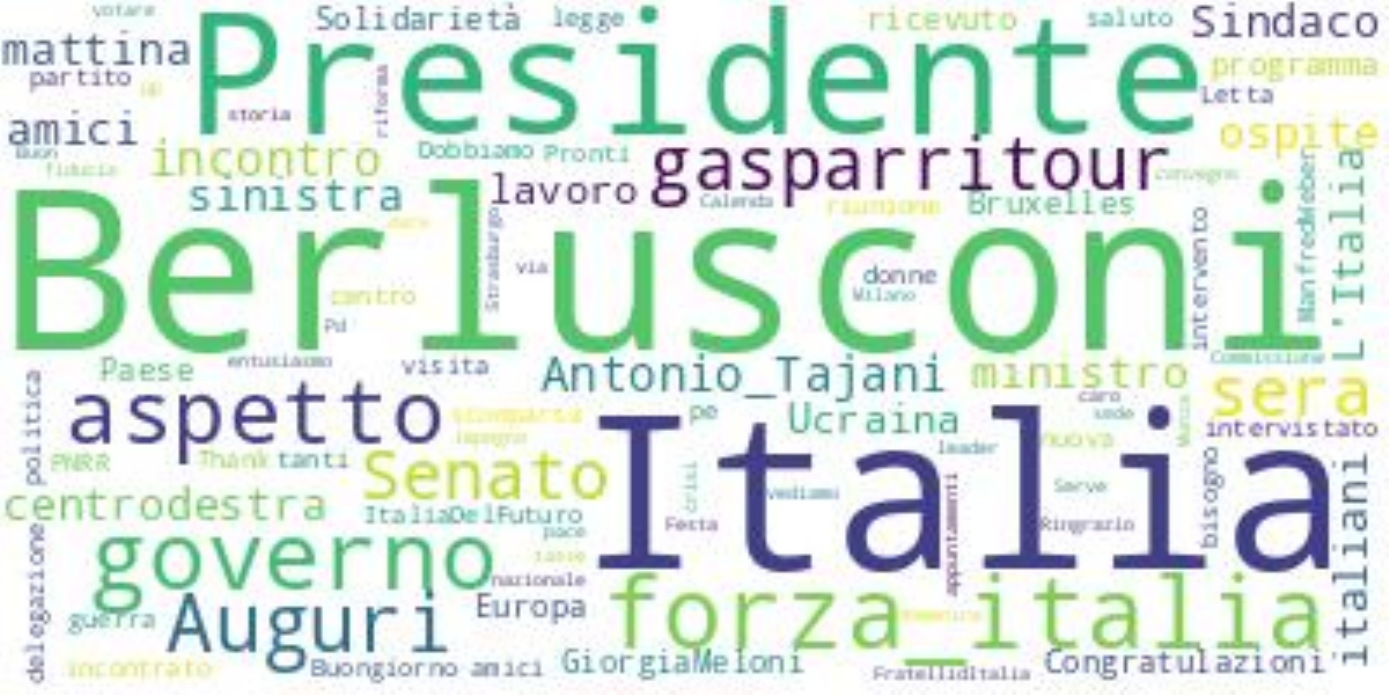} 
            }
        \caption{Word Clouds for the tweets of parties captured.}
        \label{fig:Word Clouds Parties}
\end{figure}

Our objective in this section is to answer the question, ``How did Italian politicians discuss the war?''.
After the creation of the datasets $D_1, \dots, D_6$, we cleaned each tweet by (i) removing emojis with the tool \texttt{clean-text}~\cite{Cleaner}, (ii) removing the links, and (iii) removing stop words~\cite{stopwords}. Figure~\ref{fig:Word Clouds Parties} shows the Word Clouds for each party.\footnote{We computed the word clouds using WordCloud Python Library~\cite{wordcloud}} 

The first row contains the Word Clouds associated with the parties belonging to the center-left coalition: PD focuses mostly on ``lavoro" (``job"), ``destra" (``right-wing"), and  ``Ucraina" (``Ukraine"); M5S concentrates on their own public appearance, with words like ``TV" and ``intervista" (``interview"), and its leader ``Giuseppe Conte". Finally, SiVe emphasizes their new coalition with the words ``AlleanzaVerdiSinistra" (``Green Party-Italian Left Coalition") and ``europaverde" (``Green Europe").

On the other hand, the second row is made by the parties belonging to the center-right coalition: FdI, similarly to PD, concentrates on their opposing wing with words like ``sinistra" (``left-wing") and ``governo" (``government"); Lega is vastly influenced by its leader ``Matteo Salvini" and his public appearances, indicated by words like ``TV" and ``Radio". FI rotates around its leader too, as the most commonly used words are ``Presidente" (``President") and ``Berlusconi".
Since the word clouds only provide a high-level view of the most commonly used words, we refine our analysis by inspecting the topics addressed by the parties. Indeed, political parties usually shape their campaigns by supporting or emphasizing particular themes. Thus, we extracted the topic they mainly discussed, and analyzed whether the Russo-Ukrainian war played a prominent role. To extract the topics, we started by calculating the embeddings of our tweets using the pre-trained multilingual Sentence-Bert model~\cite{reimers-2020-multilingual-sentence-bert} supporting Italian language\footnote{We used the model \texttt{distilbert-multilingual-nli-stsb-quora-ranking}.}. The corresponding tweets' embeddings (i.e., vectors of 768 dimensions) were more similar when their content was semantically closer. By leveraging this feature, we could cluster the data to find topics. First, we used UMAP algorithm~\cite{mcinnes2018umap} to decrease the vectors dimension to 5, setting n\_neighbors=15. Then, we applied the density-based HDBSCAN clustering algorithm~\cite{mcinnes2017hdbscan} to define clusters of at least 15 points, using the Excess of Mass selection method and Euclidean distance as the similarity metric. Once the clusters were defined (i.e., collections of semantically similar tweets), we extracted their most important words to manually label the corresponding topic. We calculated words' importance by using class-based TF-IDF~\cite{grootendorst2022bertopic}. In this version of the algorithm, each document corresponds to a topic (or class), i.e., the aggregation of all the tweets belonging to that topic. We can then identify the most representative words of a topic by selecting its most frequent words that are less frequent in the other topics. Table~\ref{tab:topics} shows the most discussed topics for each party, along with the percentage of tweets posted about them. For conciseness, we report only the top-7 topics for each party.

\begin{table}[ht!]
\caption{Top-7 topics and the number of tweets for each party. }
\label{tab:topics}
\resizebox{\columnwidth}{!}{%
\begin{tabular}{cc|cc|cc|cc|cc|cc}
\toprule
\multicolumn{2}{c|}{\textit{\textbf{PD}}} & \multicolumn{2}{c|}{\textit{\textbf{M5S}}} & \multicolumn{2}{c|}{\textit{\textbf{SiVe}}} & \multicolumn{2}{c|}{\textit{\textbf{FdI}}} & \multicolumn{2}{c|}{\textit{\textbf{Lega}}} & \multicolumn{2}{c}{\textit{\textbf{FI}}} \\
\textit{\textbf{\%}} & \textit{\textbf{Topic}} & \textit{\textbf{\%}} & \textit{\textbf{Topic}} & \textit{\textbf{\%}} & \textit{\textbf{Topic}} & \textit{\textbf{\%}} & \textit{\textbf{Topic}} & \textit{\textbf{\%}} & \textit{\textbf{Topic}} & \textit{\textbf{\%}} & \textit{\textbf{Topic}} \\
\midrule
24.25 & RU-UA War & 16.42 & Italy & 80.10 & Vote Left & 24.57 & Italy & 26.90 & Italy & 88.97 & Berlusconi \\
14.96 & Salary & 14.54 & Energy & 12.34 & Do & 16.28 & Vote & 17.45 & Energy & 5.10 & RU-UA War \\
10.87 & Truth & 11.12 & RU-UA War & 3.04 & RU-UA War & 12.82 & Meloni & 10.84 & RU-UA War & 1.23 & Agenda \\
10.16 & Italy & 10.35 & Mafia & 1.81 & Education & 10.27 & Do & 8.71 & Immigrants & 1.06 & Pandemic \\
8.74 & Europe & 9.15 & Salary & 0.64 & Military Exp. & 8.49 & RU-UA War & 7.15 & Taxes & 0.90 & Italy \\
7.48 & Vote & 8.81 & Agenda & 0.48 & Iran Women & 8.34 & Taxes & 6.53 & Rome & 0.85 & Foreign wars \\
6.85 & Fascism & 7.96 & Courage & 0.48 & Climate & 6.05 & Energy & 6.24 & Vote & 0.59 & Europe \\\bottomrule
\end{tabular}%
}
\end{table}

It immediately stands out that the Russo-Ukrainian conflict was a prominent topic for each party. Particularly, the topic placed in the first three positions for five out of six parties. PD mentioned the conflict the most, while FDI was the least.
By inspecting the most important words for the topic, we find the words ``sanctions'' to appear frequently for PD, M5S, Lega, and FI, ``weapons'' for PD and SiVe, and ``solidarity'' for M5S and FDI.  
In any case, this topic appears to have a similar impact on other ``internal'' matters like taxes, migrants, or energy. Only SiVe and FI show a heavily unbalanced topic frequency. In both of these cases, however, the war played a prominent role. To conclude, all major Italian parties discussed and included the war in their campaigning. 


\subsection{Temporal Analysis of Russo-Ukrainian Discussions}
\label{sec:temporal-rus}

\begin{figure}[ht!]
        \centering
            
            {
                \label{Legend_Trends}
                \includegraphics[width=.55\textwidth]{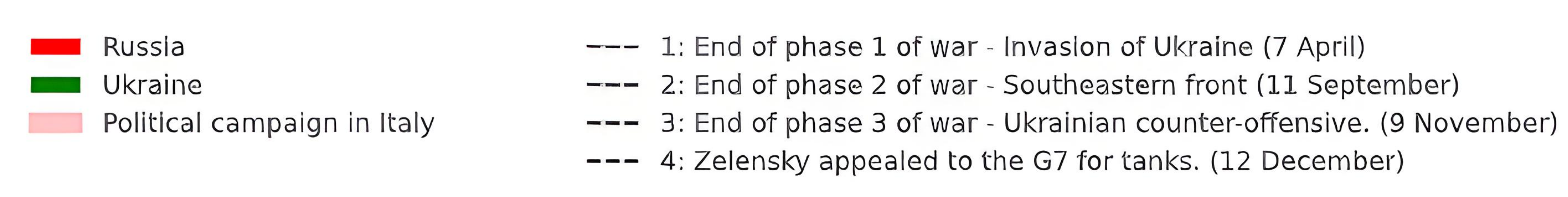}
            }\\
            \subfigure[PD] 
            {
                \label{PD_trends}
                \includegraphics[width=.47\textwidth]{Figures/Analysis/Trends/Trend_PD.pdf} 
            } 
            \subfigure[M5S] 
            {
                \label{M5S_trends}
                \includegraphics[width=.47\textwidth]{Figures/Analysis/Trends/Trend_M5S.pdf}
            }\\
            \vspace{-1em}
            \subfigure[SiVe] 
            {
                \label{SiVe_trends}
                \includegraphics[width=.47\textwidth]{Figures/Analysis/Trends/Trend_SiVe.pdf} 
            } 
            \subfigure[FdI] 
            {
                \label{FdI_trends}
                \includegraphics[width=.47\textwidth]{Figures/Analysis/Trends/Trend_FdI.pdf} 
            }\\
            \vspace{-1em}
            \subfigure[Lega] 
            {
                \label{Lega_trends}
                \includegraphics[width=.47\textwidth]{Figures/Analysis/Trends/Trend_Lega.pdf} 
            }
            \subfigure[FI] 
            {
                \label{FI_trends}
                \includegraphics[width=.47\textwidth]{Figures/Analysis/Trends/Trend_FI.pdf} 
            }
        \caption{Temporal trends for the war-related tweets, 15 days aggregation.}
        \label{fig:Trends Parties}
\end{figure}

We noted that each party included the Russo-Ukrainian war in their political campaigns. However, it is important to understand when the parties discussed it the most. We could expect, for instance, high frequencies at the beginning of the war or near the elections. In such a sense, a temporal analysis can help us understand which parties 
concentrated their whole campaigns on the war or only referred to it in crucial moments to express solidarity. 
To this aim, we created stack plots to inspect the temporal references to ``Ukraine'' and ``Russia'' during the year. Specifically, we computed the frequency of tweets related to Ukraine and Russia using a bag of words approach, i.e., by counting the number of occurrences of Ukraine/Russia-related words, such as ``Ukrainian'', ``Zelensky'' or ``Russian'', ``Putin''.
The results are presented in Figure~\ref{fig:Trends Parties}. For clarity, we also reported four major events during the conflict, such as the three main phases described in~\cite{War_Timeline} and~\cite{ISW}.

All parties discussed the Russo-Ukrainian war mostly between the beginning and end of \texttt{phase 1}. Particularly, PD shows the most active involvement, which is in accordance with Table~\ref{tab:topics}, while FI displays the highest number of tweets at the end of \texttt{phase 1}. Over the year, all parties gradually decreased their discussion of the topic, except for PD, Lega, and FI, which devoted a significant portion of their campaign propaganda. Interestingly, while Russia and Ukraine-related words were balanced initially, these parties focused most on Russia-related words during the campaign, showing a condemnation attitude rather than solidarity, as confirmed by manual inspection. The remaining parties did not accentuate the topic during the campaign, except near the end of phase 2. 

Following the election, which saw the center-right coalition led by FdI winning, there was a noticeable decline in the number of tweets related to the war from most political parties. In contrast, FdI and FI continued to post about the war, sometimes with increasing activity during \texttt{phase 3} and \texttt{phase 4}. 
In these cases, the focus seems to have switched to Ukraine rather than Russia, probably reflecting the evolution of the conflict.
These considerations suggest that while the Russian-Ukrainian war may no longer be a trending topic among most political parties, it remained quite an important issue for FdI and FI, who continue supporting Ukraine in their political messages~\cite{corriere2023schieramenti}.
\section{Bots Influence Analysis}
\label{sec:bots}
In the previous section, we highlighted that the Russo-Ukrainian conflict played a major role during the 2022 Italian General Elections. We now explore how many bots participated in the political discussions (Section~\ref{ssec:bots_presence}), 
whether bots manipulated or distorted the discussions of the Russo-Ukrainian conflict, (Section~\ref{ssec:bots_influence}), and whether they influenced real users or simply followed the flow of the conversation (Section~\ref{ssec:temporal_infl}).

\subsection{Bots Presence Analysis}
\label{ssec:bots_presence}
To evaluate the bots' influence on elections, we retrieved all replies under the posts of each party's secretary during the last month of elections, between 23 August and 23 September 2022. To detect bots among the commenters, similar to previous works on Italian tweets~\cite{martini2021bot,mattei2021italian}, we employed Botometer~\cite{yang2022botometer}, a widespread ML-based tool~\cite{shevtsov2022identification,lorenzo2022sociodemographics} that distinguishes between legitimate users and bots. 
Among the metrics, Botometer returns, for each checked account, the following scores:
\begin{enumerate}
    \item[\textbullet]\texttt{overall raw score}: score in [0, 1] determining whether an account is a bot; 
    \item[\textbullet]\texttt{cap}: (Complete Automation) Probability in [0, 1] that an account with that score or greater is a bot. In other words, it expresses the prediction's confidence.
    
\end{enumerate}
A classic approach to classify a bot takes the \texttt{overall raw score} and compares it to a fixed threshold (e.g., $>$ 0.50 classified as a bot, $\le$ 0.50 classified as human). Instead, for each user, we labeled as bot those with \texttt{overall raw score} $>$ \texttt{cap}, with \texttt{cap} $>$ 0.80 . By doing so, we adopted a dynamic and more accurate threshold than the classic approach, reducing the number of false positives. This method was confirmed by parsing several accounts manually, and among them, users with a high CAP (i.e., above 0.80) value were always classified as bots.
Table~\ref{tab:Table_Bots_Percentages} reports the number of unique accounts labeled as bots that replied under the party's secretary. On average, we found $\sim12\%$ of bots replying to each secretary, with Meloni showing the higher percentage of bots (15.08\%) and Fratoianni the lowest (9.61\%).
    
\begin{table}[ht!]
\caption{Percentages of bots and non-bots for each profile.}
\label{tab:Table_Bots_Percentages}
\centering
\scriptsize
\begin{tabular}{lccc}
            \toprule
            \textbf{Profile} & \textbf{Unique Users}  & \textbf{Bots} (\%)  & \textbf{Non-bots} (\%) \\
            \midrule
            Letta & $35,571$ & $10.76$ & $89.24$\\
            Conte & $14,255$ & $12.20$ & $87.80$\\
            Fratoianni & $2,986$ & $9.61$ & $90.39$\\ 
            Meloni & $22,670$ & $15.08$ & $84.92$ \\
            Salvini & $20,159$ & $11.12$ & $88.88$\\
            Berlusconi & $9,962$ & $12.92$ & $87.08$\\
            \bottomrule
        \end{tabular}
\end{table}

We further investigate the categories of bots interacting with Twitter profiles, according to Botometer classification. In particular, bots fall into the following categories:
\begin{enumerate}
    \item[\textbullet] \textit{Financial}: bots that post using cashtags;
    \item[\textbullet] \textit{Fake-follower}: bots purchased to increase follower counts;
    \item[\textbullet] \textit{Spammer}: accounts labeled as spambots from several datasets;
    \item[\textbullet] \textit{Self-declared}: known bots listed on botwiki.org;
    \item[\textbullet] \textit{Astroturf}: accounts that primarily focus on influencing public opinion, often being part of a network;
    \item[\textbullet] \textit{Other}: miscellaneous bots.
\end{enumerate}
 Given that Botometer's response includes a percentage indicating the likelihood of an account belonging to each category, a bot was assigned to the category with the greatest likelihood. The final cumulative results for each politician are presented in Table~\ref{tab:Table_Bots_Qualities}.
\begin{table}[htbp]
\caption{Categories of bots distribution replying to the tweets of the leaders.}
\label{tab:Table_Bots_Qualities}
\centering
\tiny
\begin{tabular}{lccccccc}
            \toprule
            \textbf{\textit{Profile}} & \textbf{\textit{Number of Bots}}& \textbf{\textit{Financial}} (\%)  & \textbf{\textit{Fake-followers}} (\%)  & \textbf{\textit{Spammers}} (\%) & \textbf{\textit{Self-declared}} (\%) & \textbf{\textit{Astroturf}} (\%) & \textbf{\textit{Other}} (\%)\\
            \midrule
            Letta & $3828$ & $0.06$ & $25.33$ & $0.15$ & $33.07$ & $35.83$ & $5.56$\\
            Conte & $1739$ & $0.08$ & $33.87$ & $0.08$ & $31.27$ & $32.04$ & $2.67$\\
            Fratoianni & $287$ & $0.00$ & $19.44$ & $0.00$ & $42.78$ & $31.67$ & $6.11$\\
            Meloni & $3418$ & $0.04$ & $30.03$ & $0.15$ & $33.53$ & $31.50$ & $4.75$\\
            Salvini & $2242$ & $0.06$ & $39.35$ & $0.11$ & $27.97$ & $27.69$ & $4.83$\\
            Berlusconi & $1287$ & $0.44$ & $26.40$ & $0.00$ & $31.79$ & $34.43$ & $6.93$\\
            \bottomrule
        \end{tabular}
\end{table}

 A significant proportion of counterfeit profiles engaged with political figures fall under the categories of ``fake\_followers'' and ``astroturf''. This result confirms that most analyzed bots aim to influence or manipulate public opinion. Another notable percentage pertains to ``self-declared'' bots that, on the other hand, operate on the platform without any nefarious motives.
In general, the bots distribution is consistent across all profiles.

\begin{figure}[ht!]
        \centering
            \subfigure[Center-Left coalition] 
            {
                \label{fig:venn_diagram_leftWing}
                \includegraphics[width=.3\textwidth]{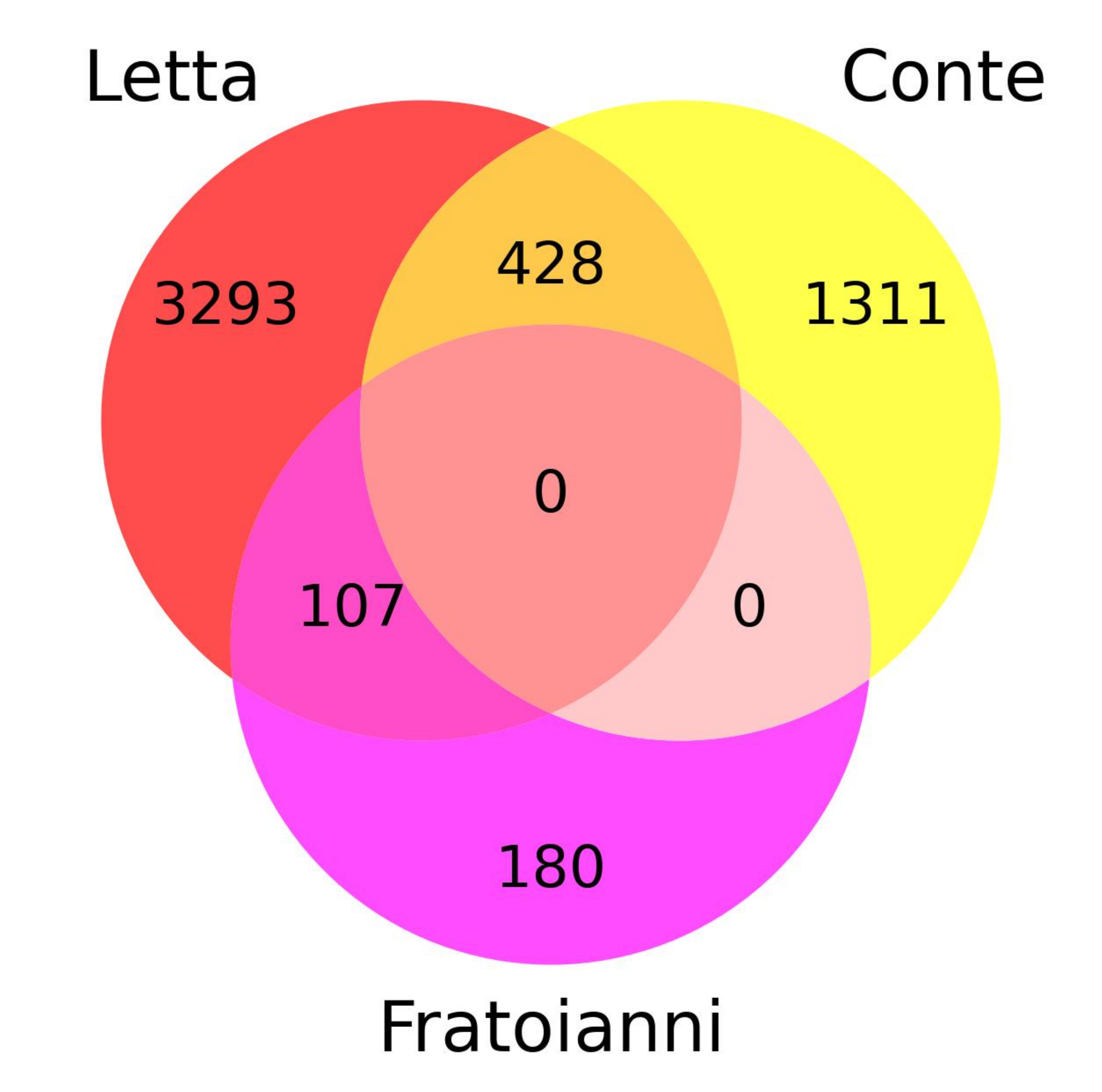} 
            } 
            \hspace{2cm}
            \subfigure[Center-Right coalition] 
            {
                \label{fig:venn_diagram_rightWing}
                \includegraphics[width=.3\textwidth]{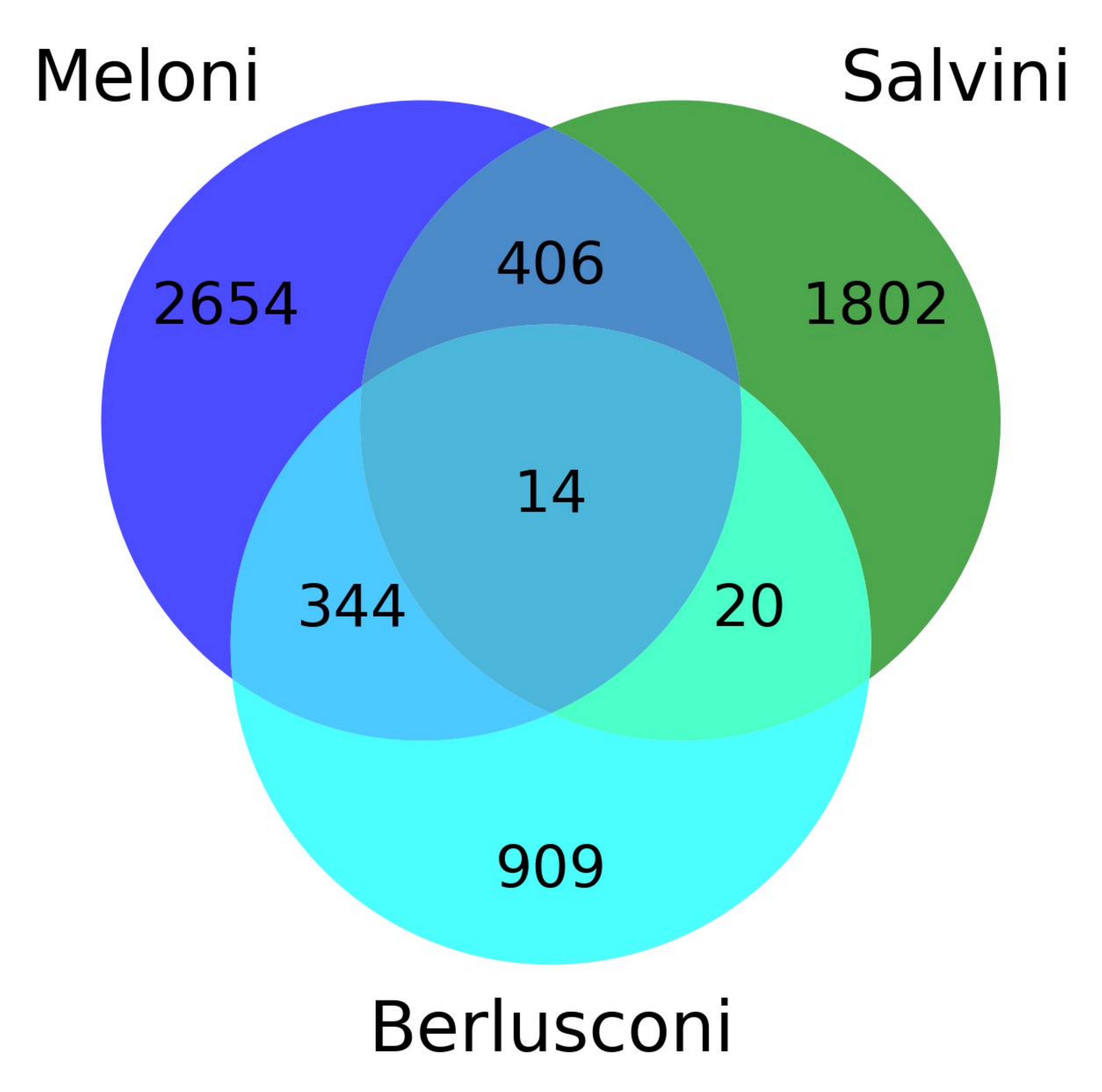} 
            }
         \caption{Number of shared bots between profiles belonging to the same coalition. Colors are representative of the parties, according to the Italian press.}
        \label{fig:venn_diagrams}
\end{figure}

We further investigate
whether bots cooperate within the two coalitions we described in Section~\ref{sec:dataset}, namely, the Center-Right coalition (Berlusconi, Meloni, and Salvini) and the Center-Left coalition (Letta, Conte, and Fratoianni). Figure~\ref{fig:venn_diagrams} shows the shared number of bots in the two coalitions.
For the Center-Left coalition, many accounts identified as bots and commenting on multiple politicians are associated with Letta and Conte, the primary figures in the ``giallo-rosso government'' mentioned earlier. Additionally, the remaining shared bots are linked to Fratoianni and, once again, Letta, the leaders of the two largest parties comprising the Center-Left coalition in the most recent elections. On the other hand, in the Center-Right coalition, there is a significantly stronger affiliation between the three profiles, as confirmed by the interrelation between the three political parties. Several bot accounts are common to two profiles, with a select few being shared by all three, suggesting a much closer connection between the coalition's parties and their ideologies.

\subsection{Bots Topics Distortion Analysis}
\label{ssec:bots_influence}
We now investigate the lexical associations between the words employed by authentic and bot users during the last month of the Italian General Elections' political campaign. In this way, we can explore and understand how bots and humans communicated regarding the Russo-Ukrainian conflict, and whether bots distorted the vision of war-related topics.
Inspired by the methodology introduced in Sartori et al.~\cite{sartori2023impact} and Tahmasbi et al.~\cite{tahmasbi2021go}, we aim to discover associations between war-related words, e.g., how frequently they appear together in a tweet.
For this purpose, we first trained a Word2Vec model~\cite{Word2Vec} on our tweets  to determine how words related to the Russian-Ukrainian war relate to each other. In this model, words with similar vectors are likelier to appear together in a tweet. Starting from the words ``Russia'', ``Ukraine'', and ``War'', we manually identified 10 frequent related words, selecting (i) institutional-related words, i.e., ``USA'', ``EU'', ``NATO'', ``Europe'', and ``Italy'';
(ii) war-related words, i.e., ``weapons'', ``conflict'', ``invasion'', ``aggression';
(iii) ``gas'', as its price rose sharply due to the conflict. 
Subsequently, we calculated the incidence matrix $M\in\mathbb{R}^{3\times10}$ for each involved party, utilizing the trained Word2Vec model. The incidence matrix $M$ can be mathematically formulated as in the Matrix~\ref{eq_incidence_matrix}.

\vspace{0.2cm}
\begin{equation}
    M=
    \begin{pmatrix}
        m_{1,1} & m_{1,2} & \dots & m_{1,9} &m_{1,10}\\
        & & & & \\
        m_{2,1} & m_{2,2} & \dots & m_{2,9} &m_{2,10}\\
        & & & & \\
        m_{3,1} & m_{3,2} & \dots & m_{3,9} &m_{3,10}
    \end{pmatrix}
    \label{eq_incidence_matrix}
\end{equation}
\vspace{0.2cm}

where $m_{ij}=$ \texttt{cosine\_similarity}$(v_{i}, w_{j})$, $i=1,2,3$ and $j=1, \dots, 10$\footnote{The \texttt{cosine-similarity} was computed according to the formula in~\cite{cosine_similarity}}. The words $v_{i}$ are the selected words \{``Russia'', ``Ukraine'', ``War''\}, while the words $w_{j}$ are the selected words \{``USA'', ``EU'', ``NATO'', ``Europe'',  ``Italy'', ``weapons'', ``conflict'',  ``invasion'', ``aggression'', ``gas''\}. If the cosine similarity was negative, we truncated it to 0. This matrix $M$ was computed for each party in two different scenarios:
\begin{enumerate}
    \item[\textbullet] A \textit{Complete} scenario, considering both replies from real and bot accounts;
    \item[\textbullet] A \textit{No Bots} scenario, considering only replies from real users.
\end{enumerate}
We fed these matrices to the Gephi Software \cite{bastian2009gephi} to construct weighted undirected graphs, which we call ``Spider Graphs'' due to their shape, and we used Force-Atlas 2~\cite{jacomy2014forceatlas2} as Layout for the rendering. In our graphs, the nodes are the words, and the edges represent the cosine similarity. According to the incidence matrix, edges exist only between the
three initial words (``Russia'', ``Ukraine'', ``War'') and the 10 selected words.
The node size reflects its degree (larger words have more connections), while the thickness of the edges reflects the similarity of the connected words (thicker edges connect more similar -- or likely to appear together -- words). Last, we applied the modularity algorithm~\cite{blondel2008fast} to build clusters of strictly connected words. 
\begin{figure}[ht!]
        \centering
        \hspace{-1.8cm}
            \subfigure[PD - Complete] 
            {
                \label{PD_SG_Complete}
                \includegraphics[height = 2.5cm]{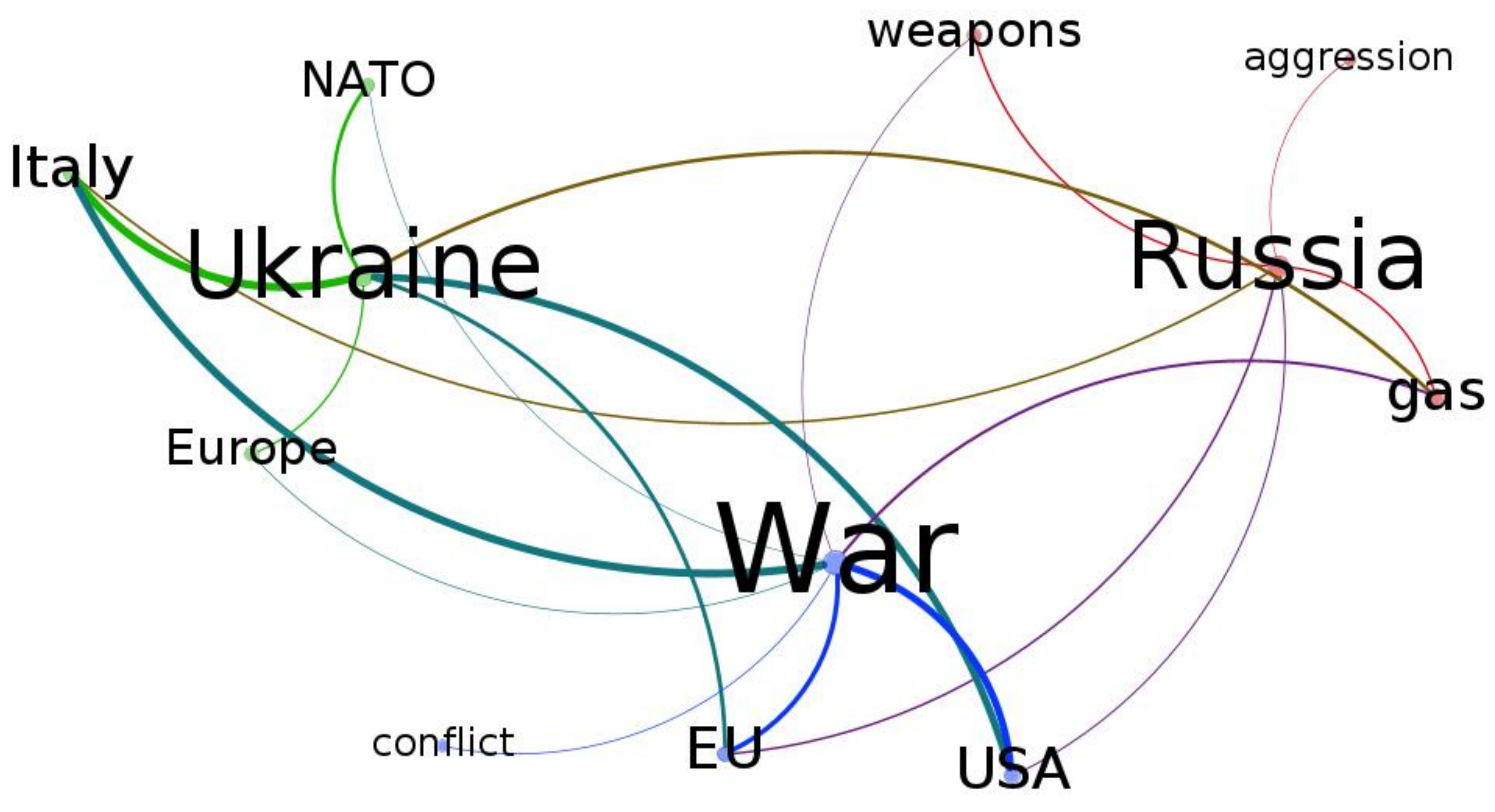} 
            } 
            \subfigure[PD - No Bots] 
            {
                \label{PD_SG_NoBots}
                \includegraphics[height = 2.5cm]{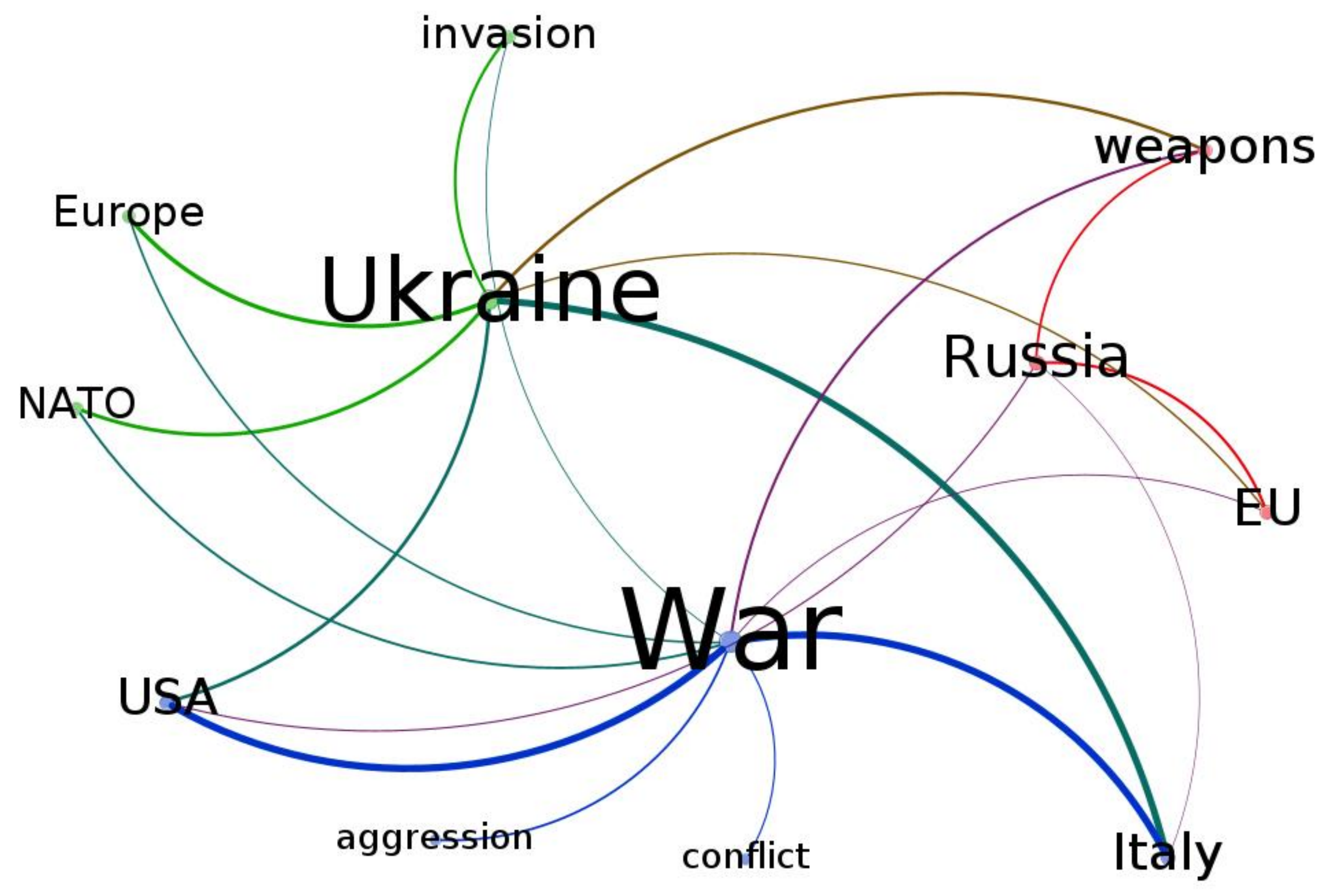} 
            }\\
            \subfigure[M5S - Complete] 
            {
                \label{M5S_SG_Complete}
                \includegraphics[height = 2.5cm]{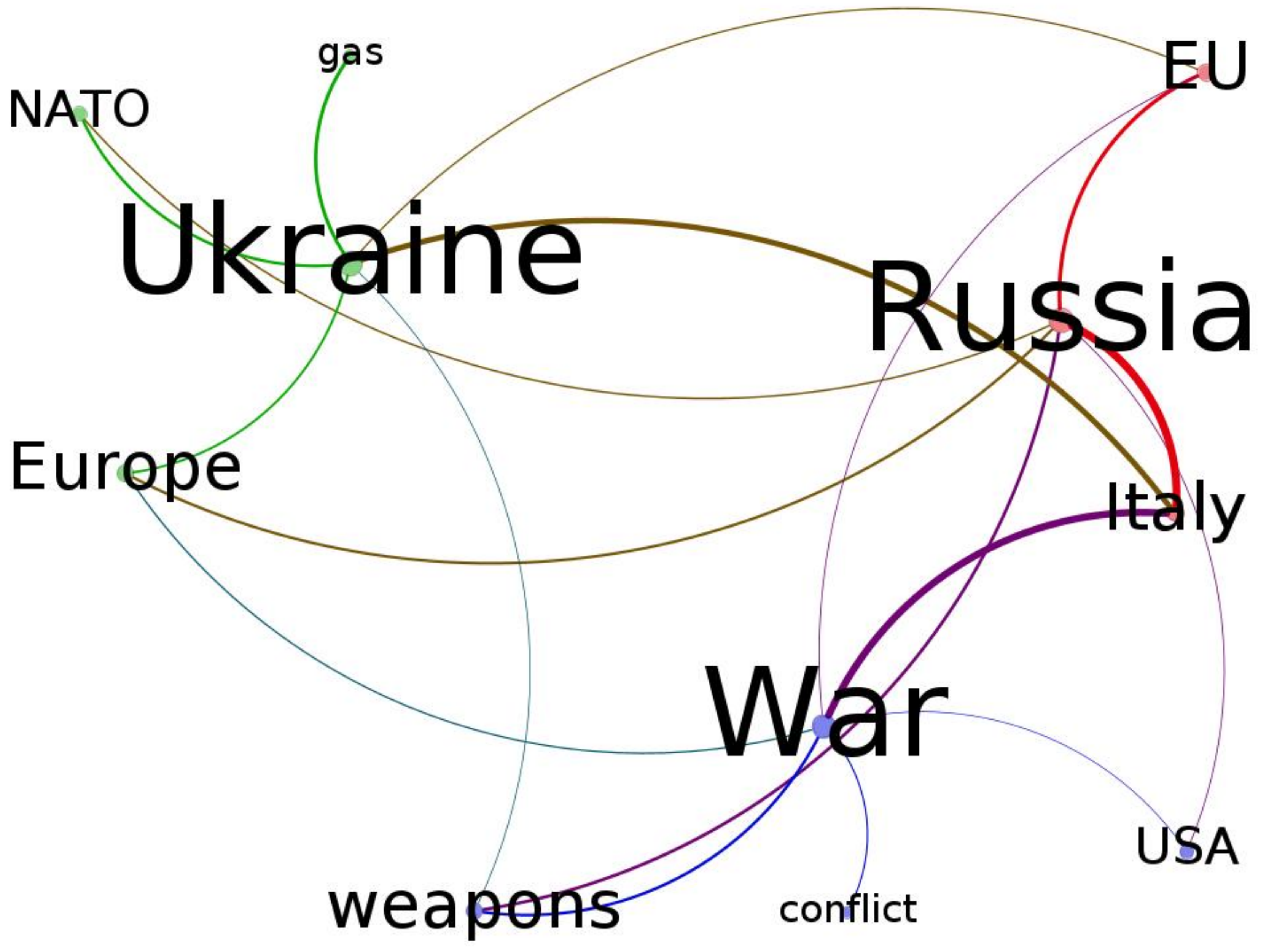}
            }
            \subfigure[M5S - No Bots] 
            {
                \label{M5S_SG_NoBots}
                \includegraphics[height = 2.5cm]{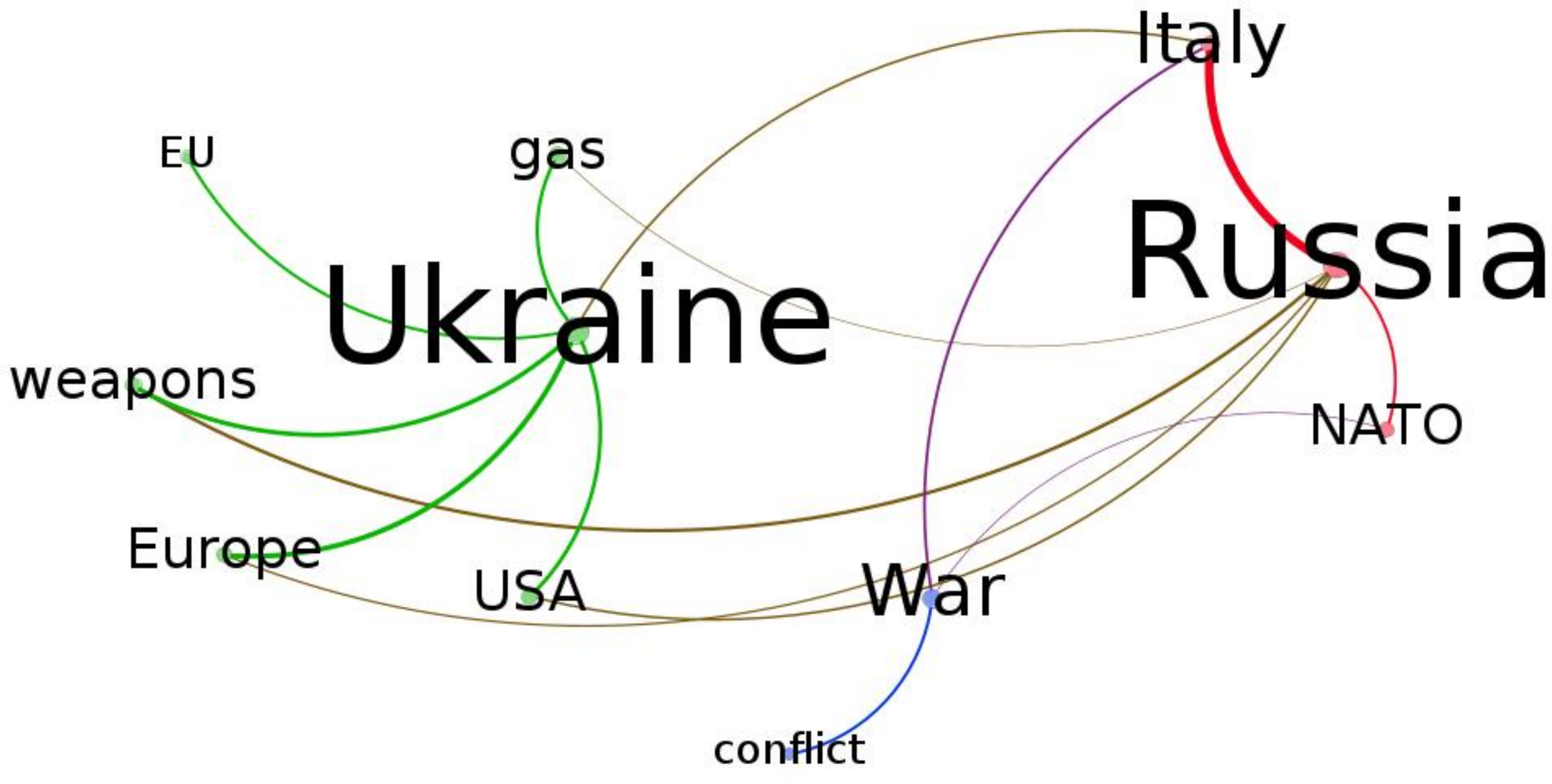}
            }\\
            \subfigure[SiVe - Complete] 
            {
                \label{SiVe_SG_Complete}
                \includegraphics[height = 2.5cm]{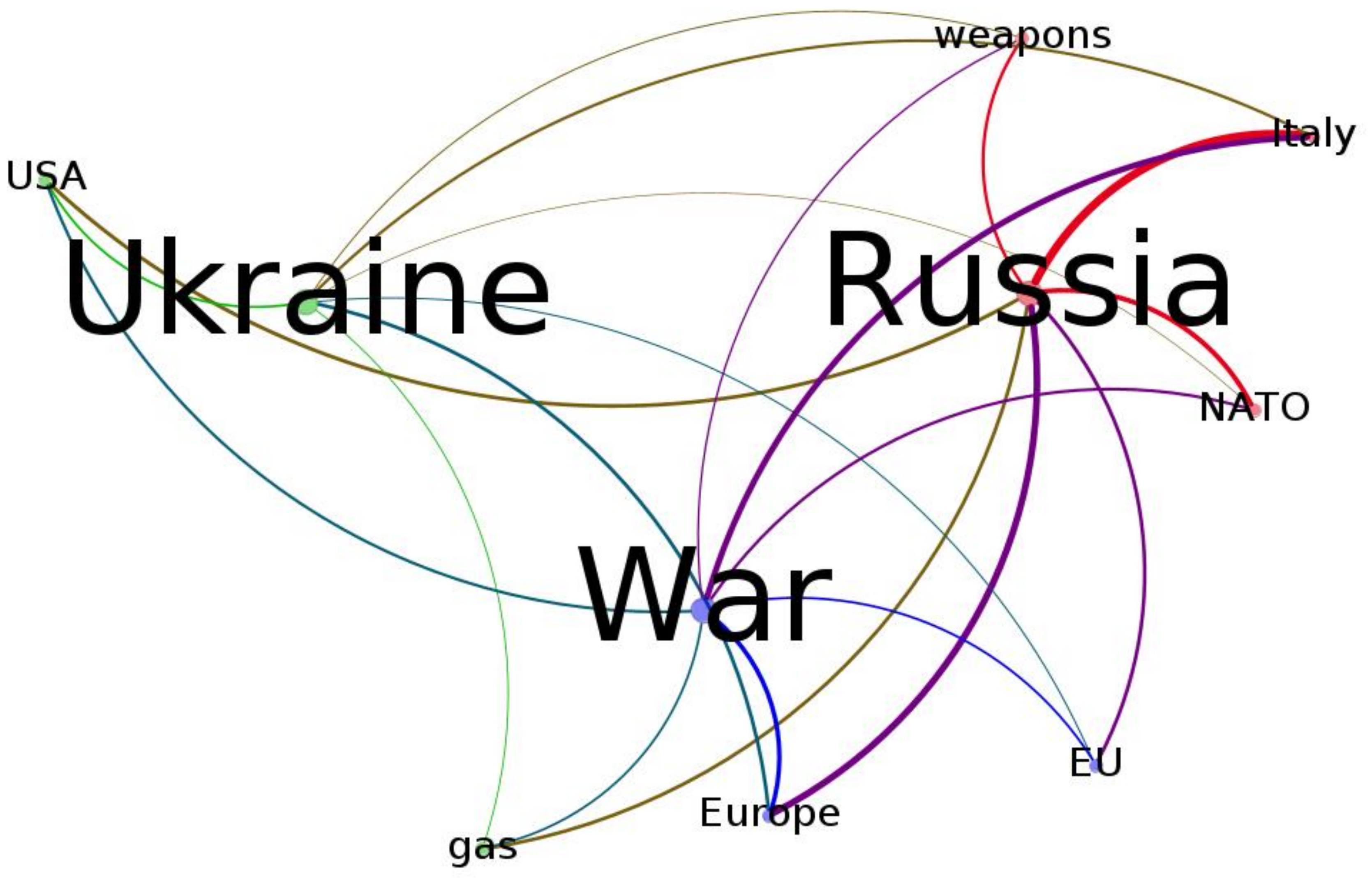} 
            } 
            \subfigure[SiVe - No Bots] 
            {
                \label{SiVe_SG_NoBots}
                \includegraphics[height = 2.5cm]{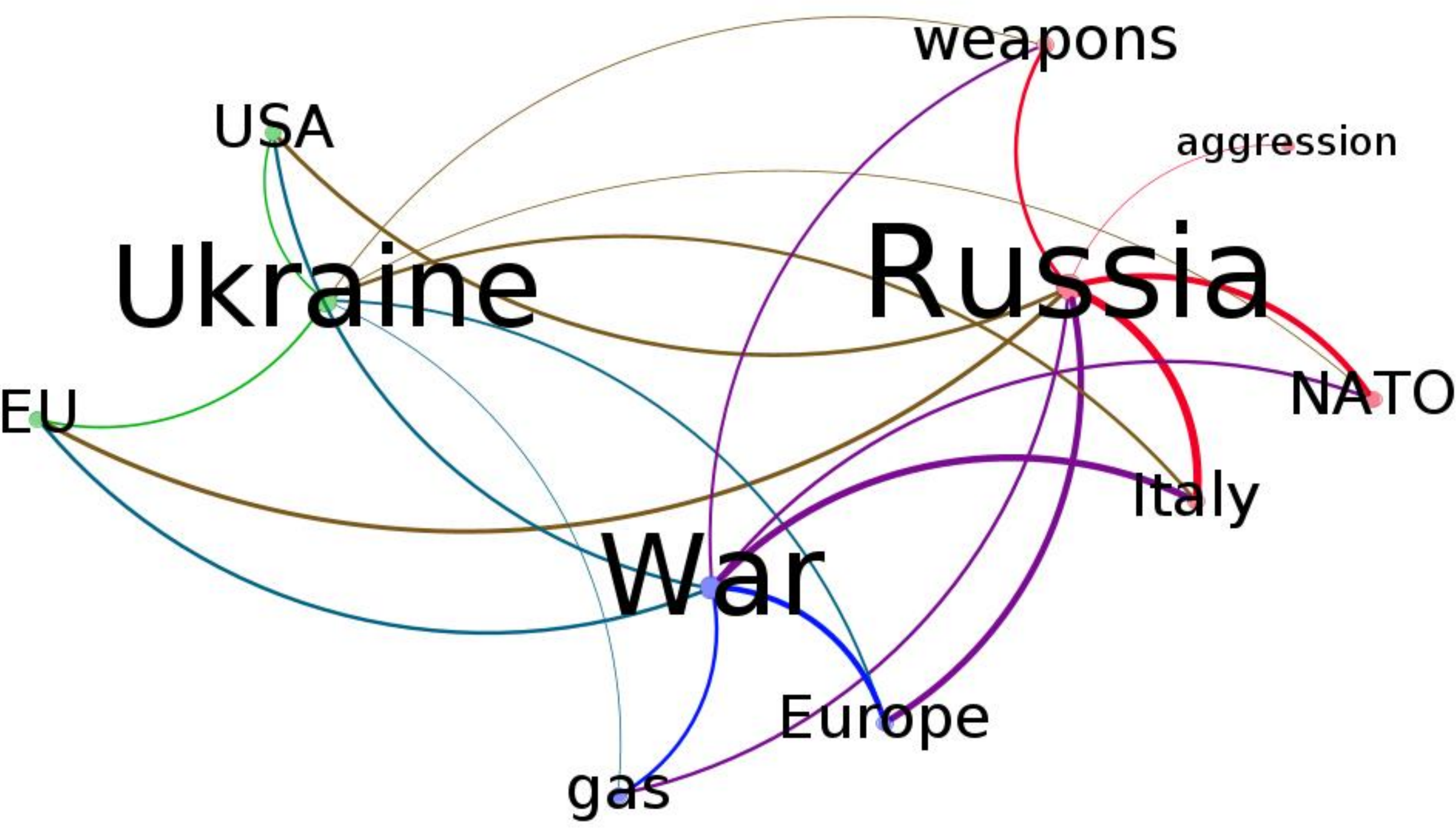} 
            }
            \caption{Comparison between ``Spider Graphs'' of the Mixed and No-Bots Scenario in the Center-Left coalition.}
        \label{fig:SpiderGraphs_CSX}
\end{figure}
\begin{figure}[ht!]
        \centering
            \subfigure[FdI - Complete] 
            {
                \label{FdI_SG_Complete}
                \includegraphics[height = 2.5cm]{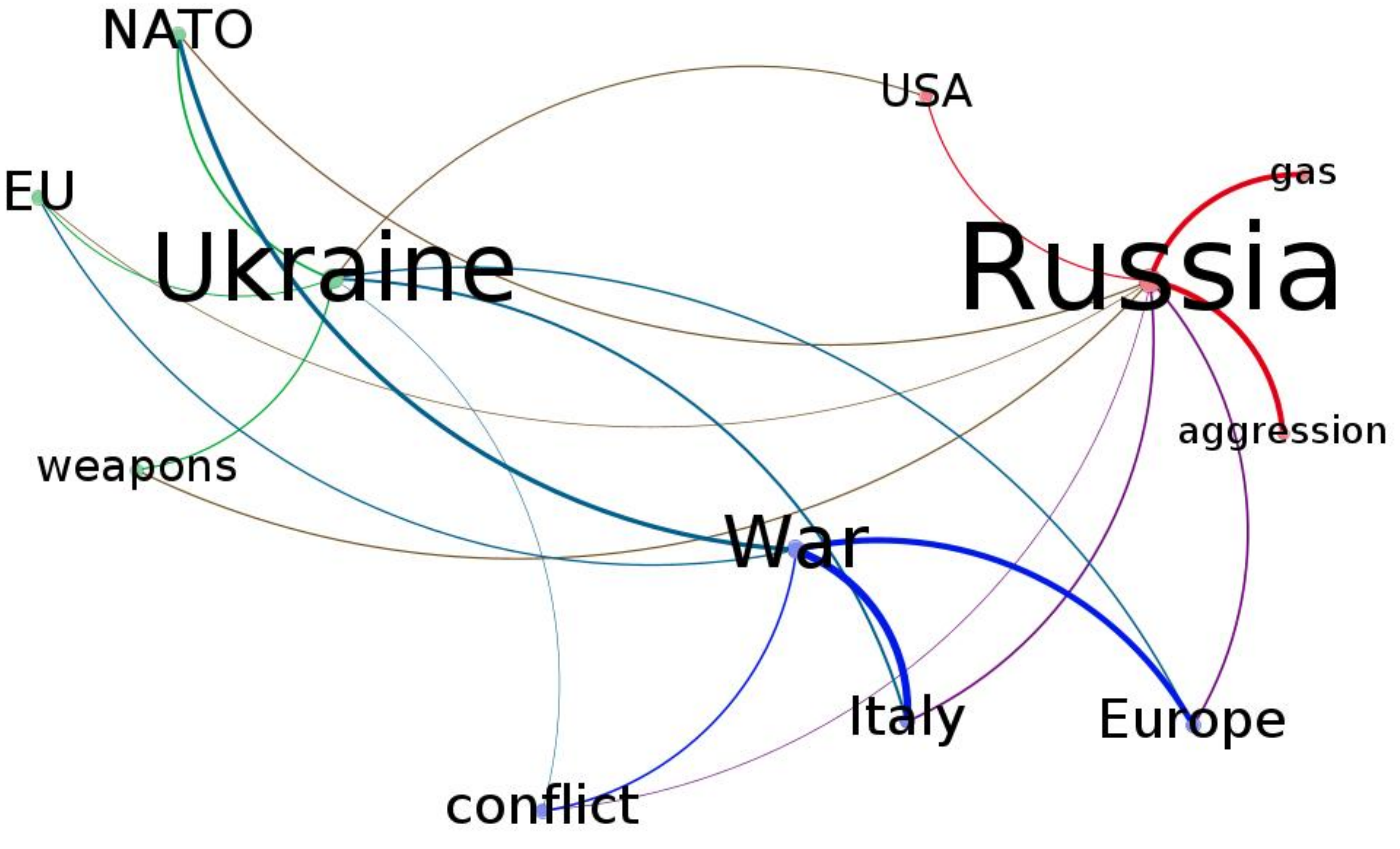} 
            }
            \subfigure[FdI - No Bots] 
            {
                \label{FdI_SG_NoBots}
                \includegraphics[height = 2.5cm]{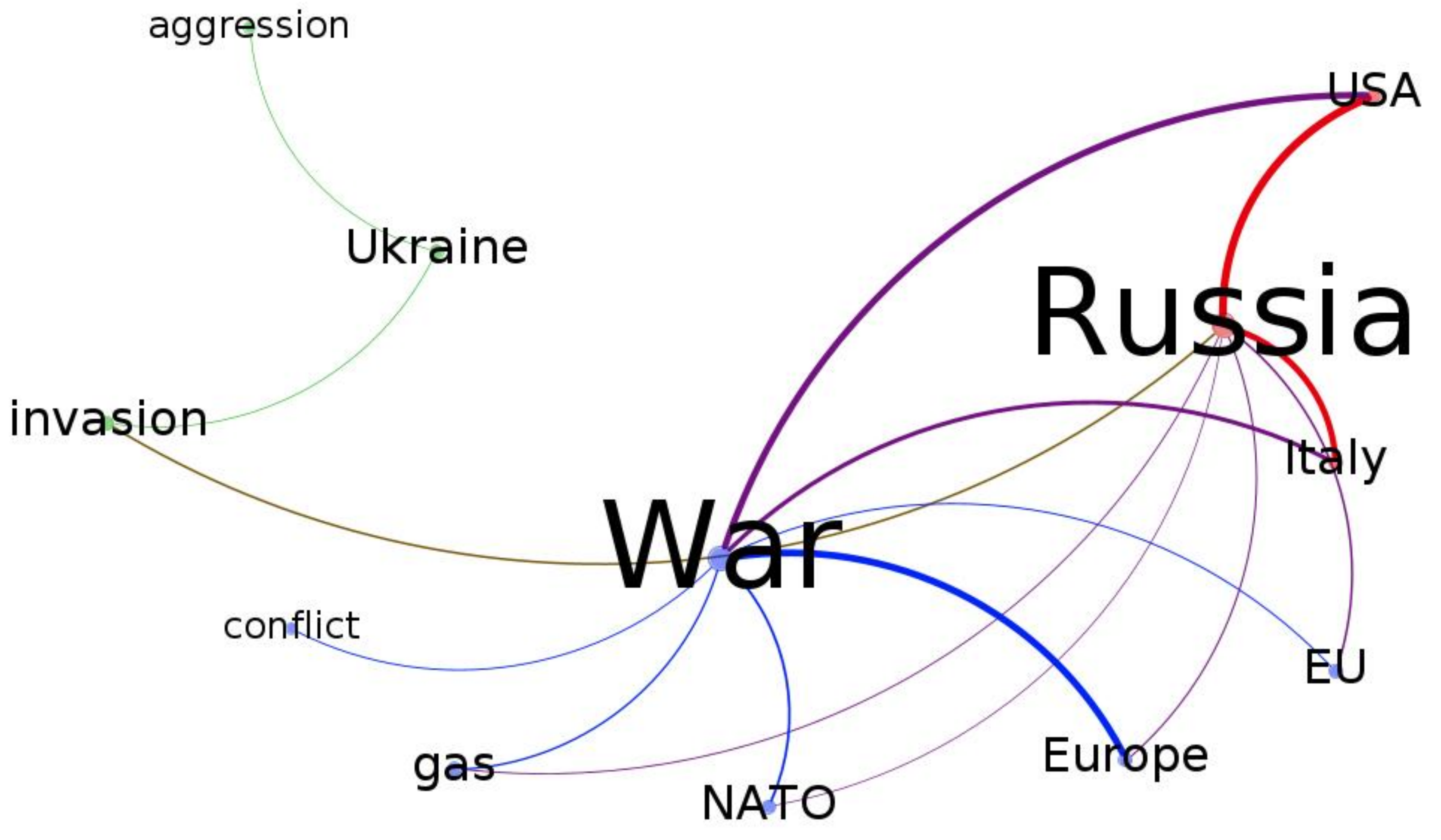} 
            }\\
            \subfigure[Lega - Complete] 
            {
                \label{Lega_SG_Complete}
                \includegraphics[height = 2.5cm]{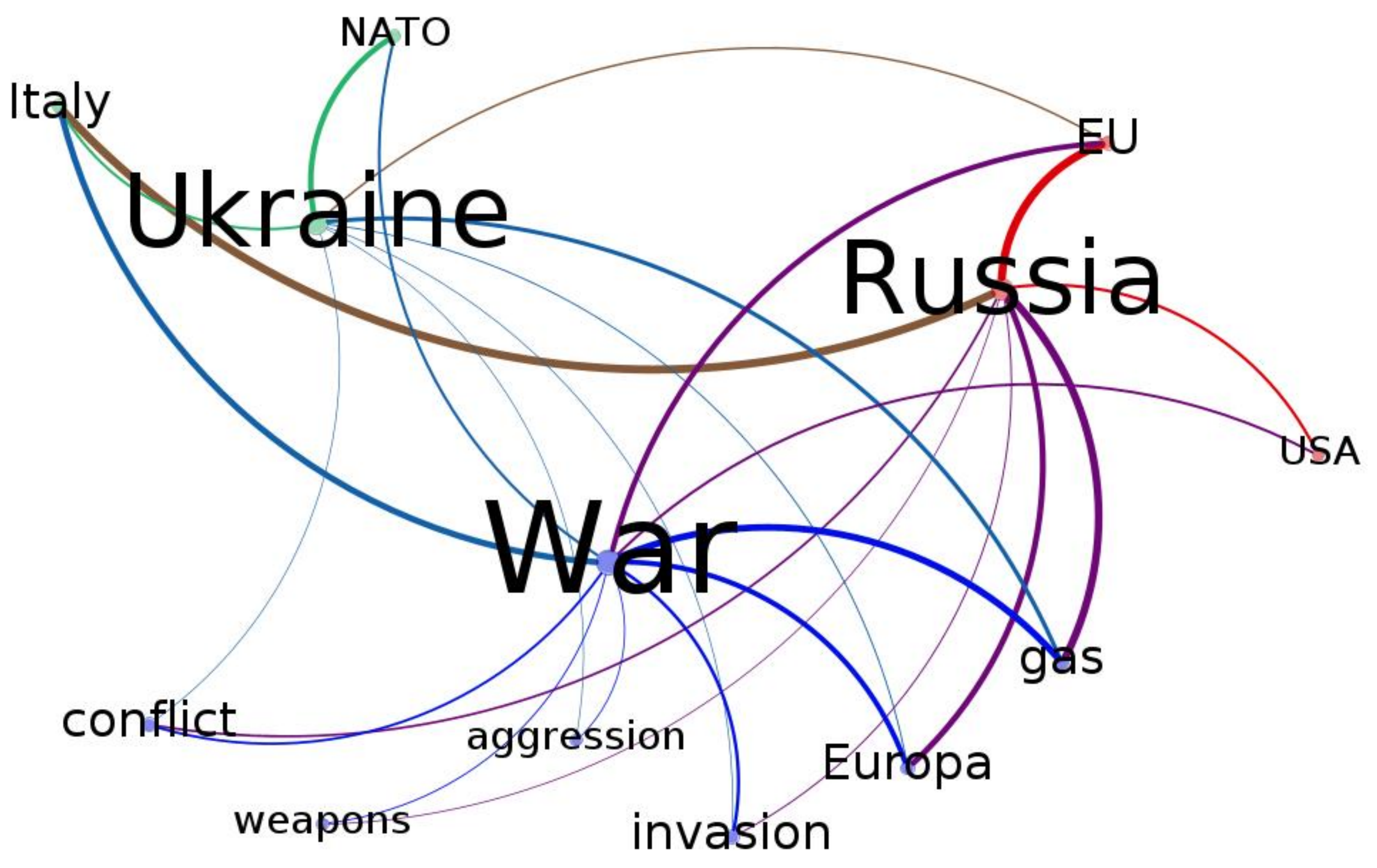} 
            }
            \subfigure[Lega - No Bots] 
            {
                \label{Lega_SG_NoBots}
                \includegraphics[height = 2.5cm]{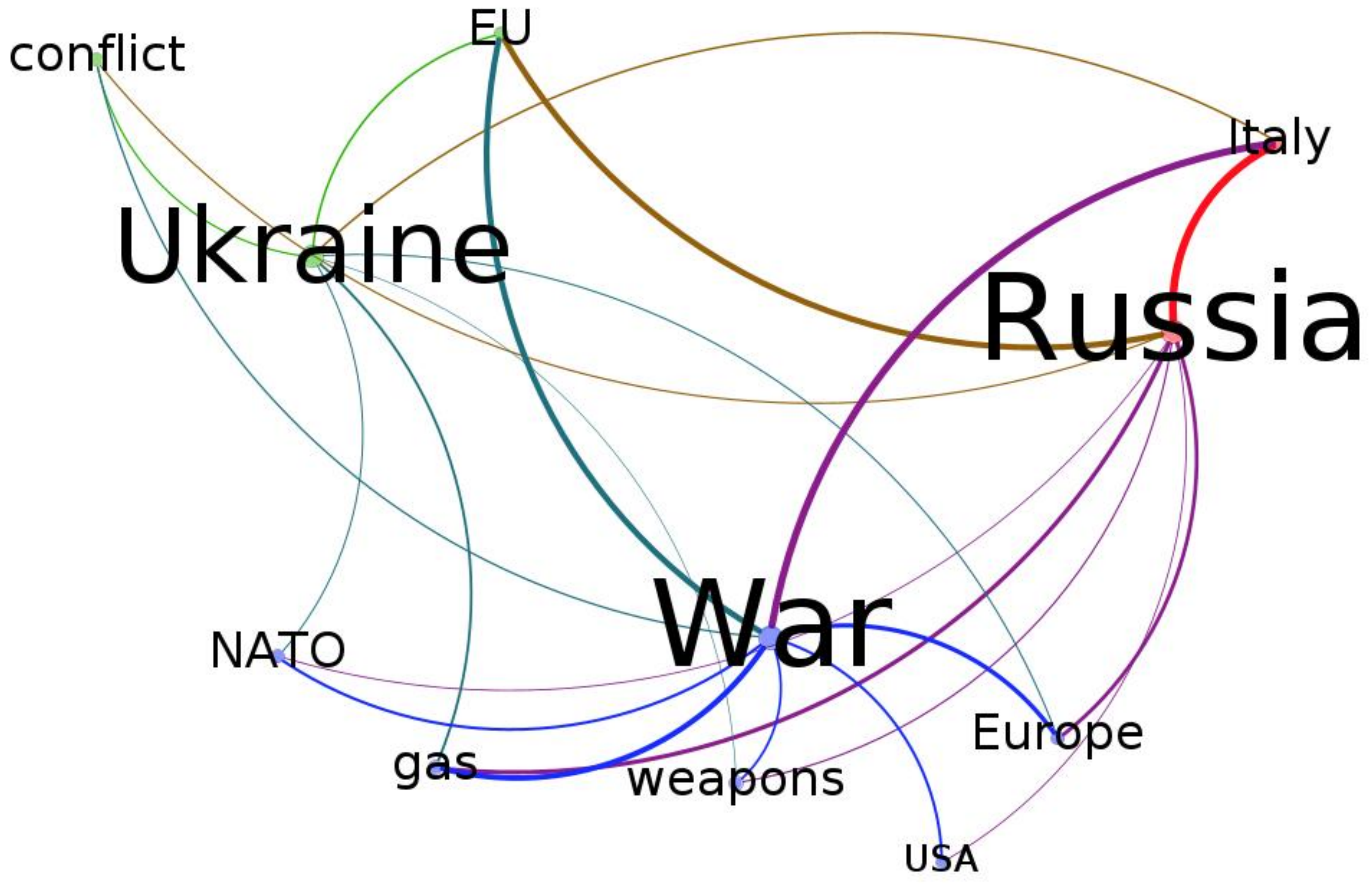} 
            }\\
            \subfigure[FI - Complete] 
            {
                \label{FI_SG_Complete}
                \includegraphics[height = 2.5cm]{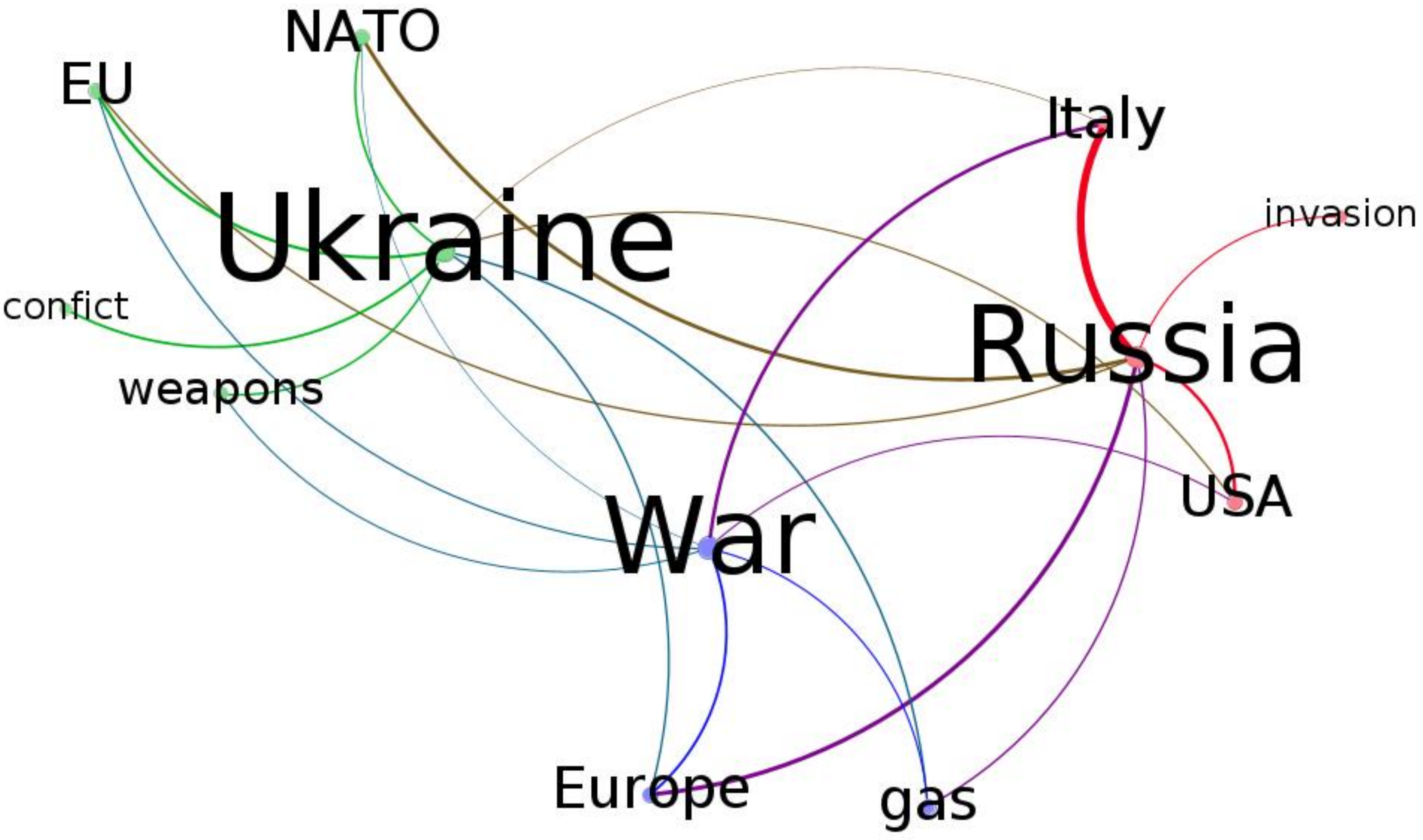} 
            }
            \subfigure[FI - No Bots] 
            {
                \label{FI_SG_NoBots}
                \includegraphics[height = 2.5cm]{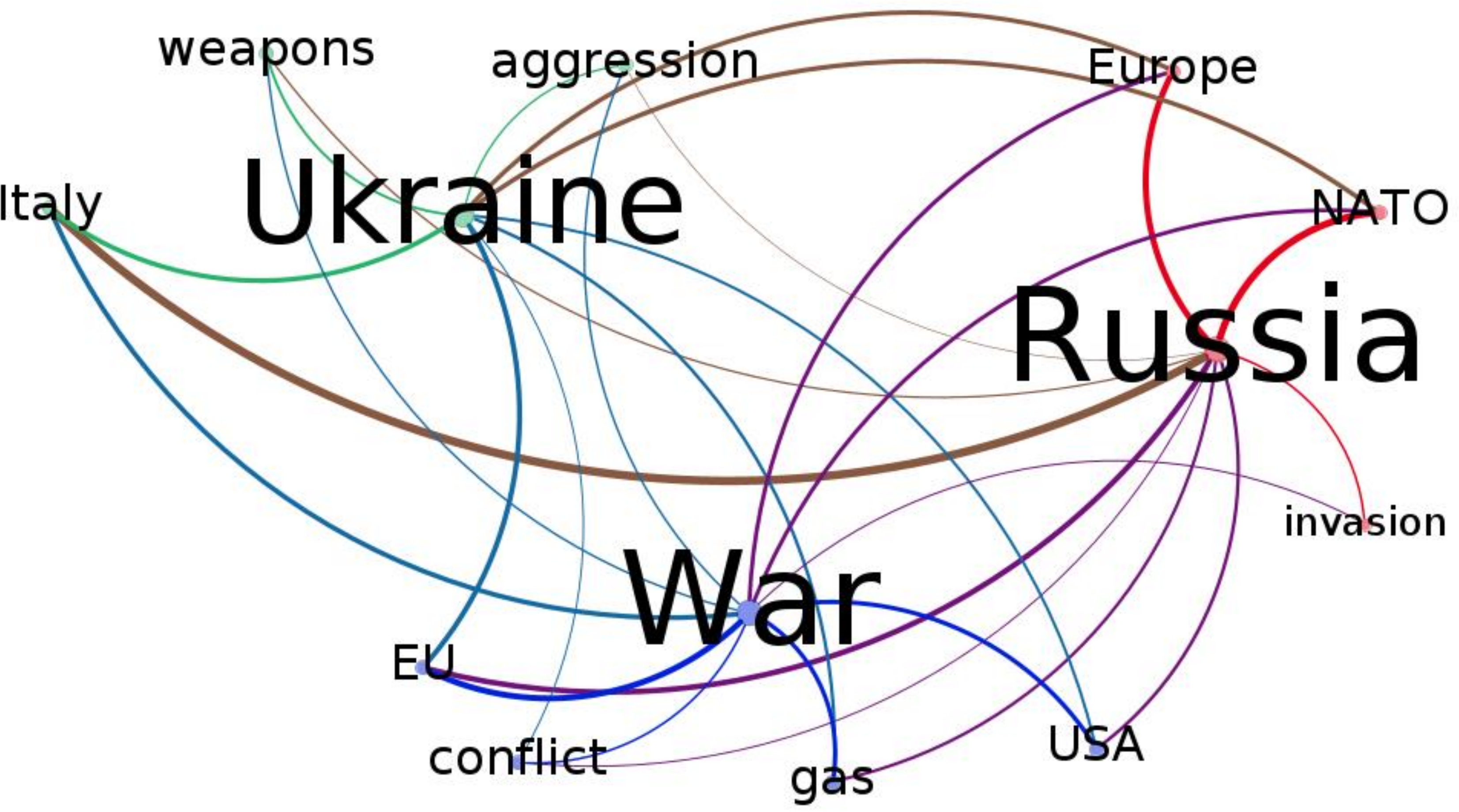} 
            }
        \caption{Comparison between ``Spider Graphs'' of the Complete and No Bots Scenario in the Center-Right coalition.}
        \label{fig:SpiderGraphs_CDX}
\end{figure}
We set the resolution parameter of the modularity algorithm to obtain three clusters: a red cluster with centroid ``Russia'', a green cluster with centroid ``Ukraine'', and a blue cluster with centroid ``War''. The remaining 10 words are then placed  by the algorithm in the closest cluster, acquiring its color. 
Edges have the color of the cluster if they are connected to their centroid, or a mixed color if they are connected to the centroid of a different cluster. For instance, the edge between a word of the ``War'' (blue) cluster and ``Russia'' (red) will be purple (blue + red). Figure~\ref{fig:SpiderGraphs_CSX} and Figure~\ref{fig:SpiderGraphs_CDX} present the spider graphs in the \textit{Complete} and \textit{No Bots} scenario of the Center-Left and Center-Right coalitions, respectively.

For the Center-Left coalition, the most significant change between the two scenarios concerns the M5S party. While the lexical similarity between ``War'' and ``conflict'' remains the same, there are no other words in the ``War'' cluster when considering the \textit{No Bots} scenario. An important constant between the two M5S graphs is the strong link between the central node ``Russia'' and ``Italy''. The graphs of PD and SiVe seem to show several differences. In the graph~\ref{PD_SG_NoBots} the word ``gas'' disappears and the cluster of ``War'' gains the word ``Italy'' from the ``Ukraine'' cluster. The word ``weapons'' is always clustered with ``Russia'' in all the graphs of PD and SiVe and the word ``Italy'' is always with Russia in M5S and SiVe. Moreover, the word ``conflict'' is present only in the graphs of PD and M5S and it is absent from the ones of SiVe. The presence in the three clusters of institutional-related words, i.e. ``NATO'', ``USA'', ``EU'' and ``Europe'', seems not to have such relevant lexical importance for the parties except for ``USA'' and ``War'' in the PD scenarios.

For the Center-Right coalition, the primary observation concerns the intensified association between the central term ``Russia'' and words that frequently pertain to institutions, such as ``Italy'', ``NATO'', and ``Europe''. Another identifiable characteristic noted by the model is the substantial presence of words within the cluster associated with the term ``War'', whereby the most frequent ones include ``gas'', ``invasion'', and ``conflict''. Within this coalition, it appears that every ``Russia'' cluster encompasses a closely related term, such as ``Italy'' or ``USA'', with a strong connection. 
The strongest differences between the two scenarios appear around the ``Ukraine'' cluster. Indeed, for all three parties, the words within the cluster differ significantly between the \textit{Complete} and \textit{No Bots} scenarios. For instance, for FdI, the ``Ukraine'' cluster goes from ``aggression'' and ``invasion'' in the \textit{No Bots} scenario to ``NATO'', ``EU'', and ``weapons'' in the \textit{Complete} scenario.  Significant differences between the scenarios also appear around the ``Russia'' cluster. Therefore, we notice how bots significantly impacted public opinion by going in the opposite direction of real users.
Considering all the graphs, we can assert that the existence of bots appears to influence the outcomes of the clustering analysis, especially for the Center-Right coalition.

\subsection{Bots Temporal Influence Analysis}
\label{ssec:temporal_infl}
\begin{figure}[ht!]
        \centering
            {
                \label{Legend_Slope}
                \includegraphics[width=.75\textwidth]{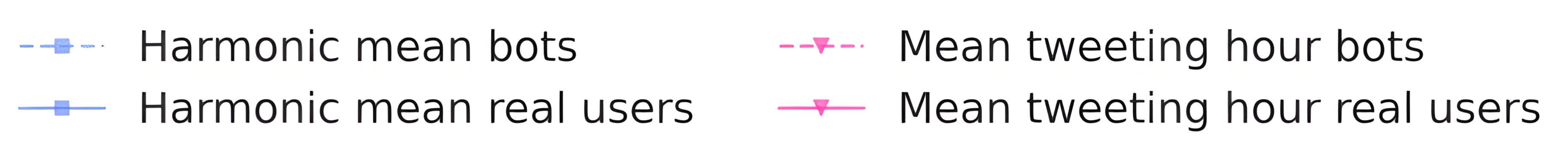} 
            }\\
            \subfigure[Letta] 
            {
                \label{Letta_Slope}
                \includegraphics[width=.47\textwidth]{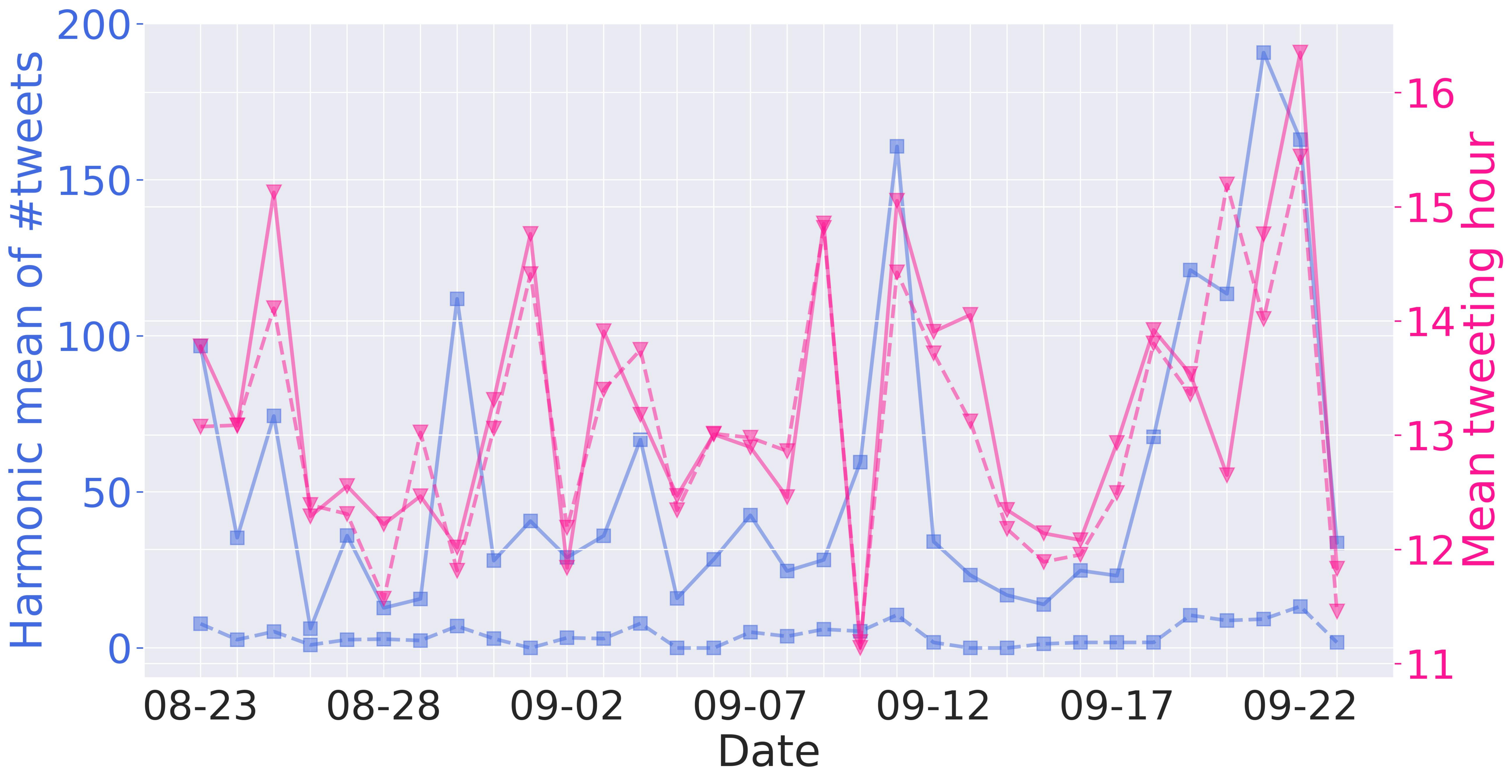} 
            } 
            \subfigure[Conte] 
            {
                \label{Conte_Slope}
                \includegraphics[width=.47\textwidth]{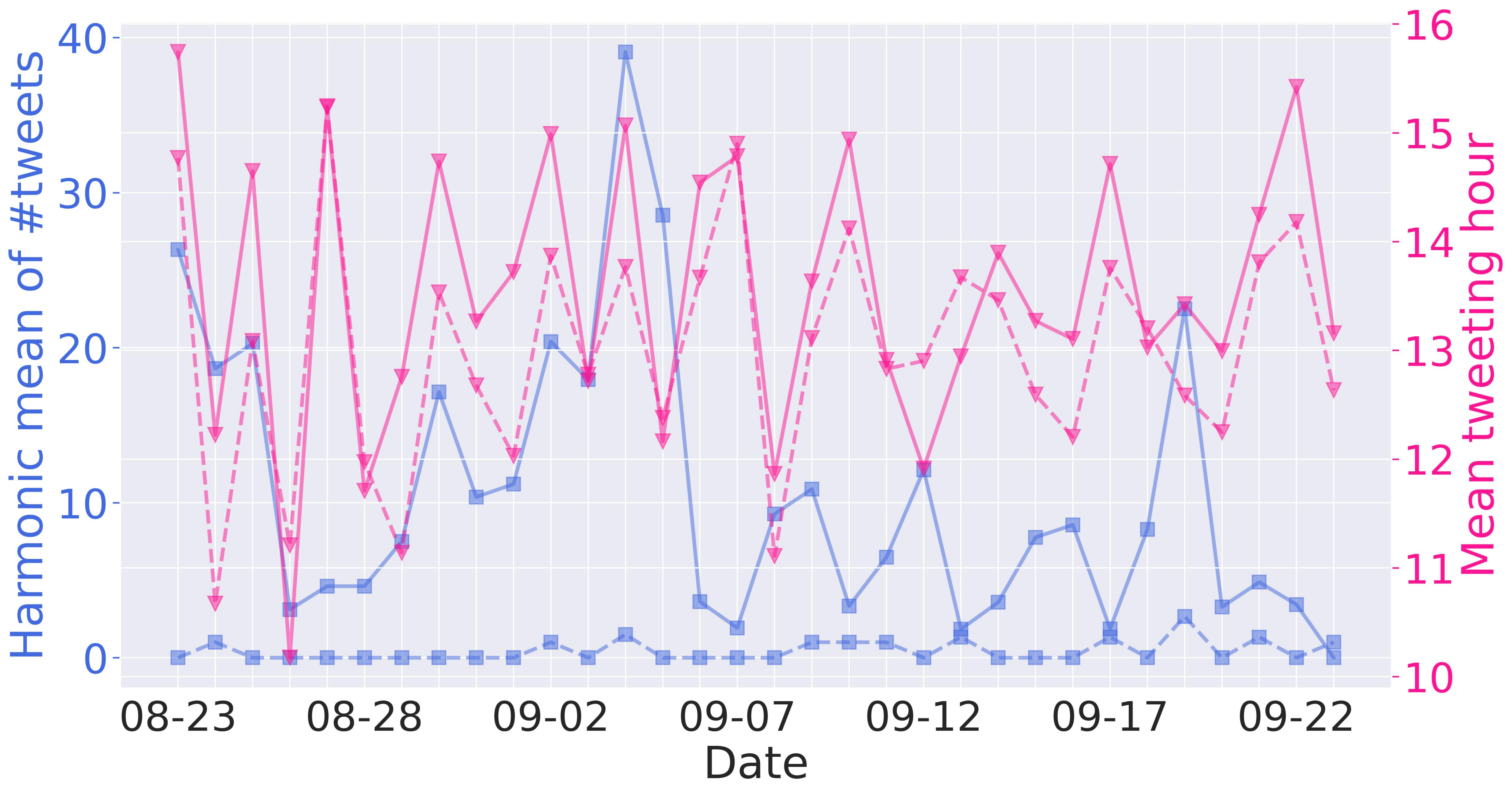} 
            }\\
            \subfigure[Fratoianni] 
            {
                \label{Fratoianni}
                \includegraphics[width=.47\textwidth]{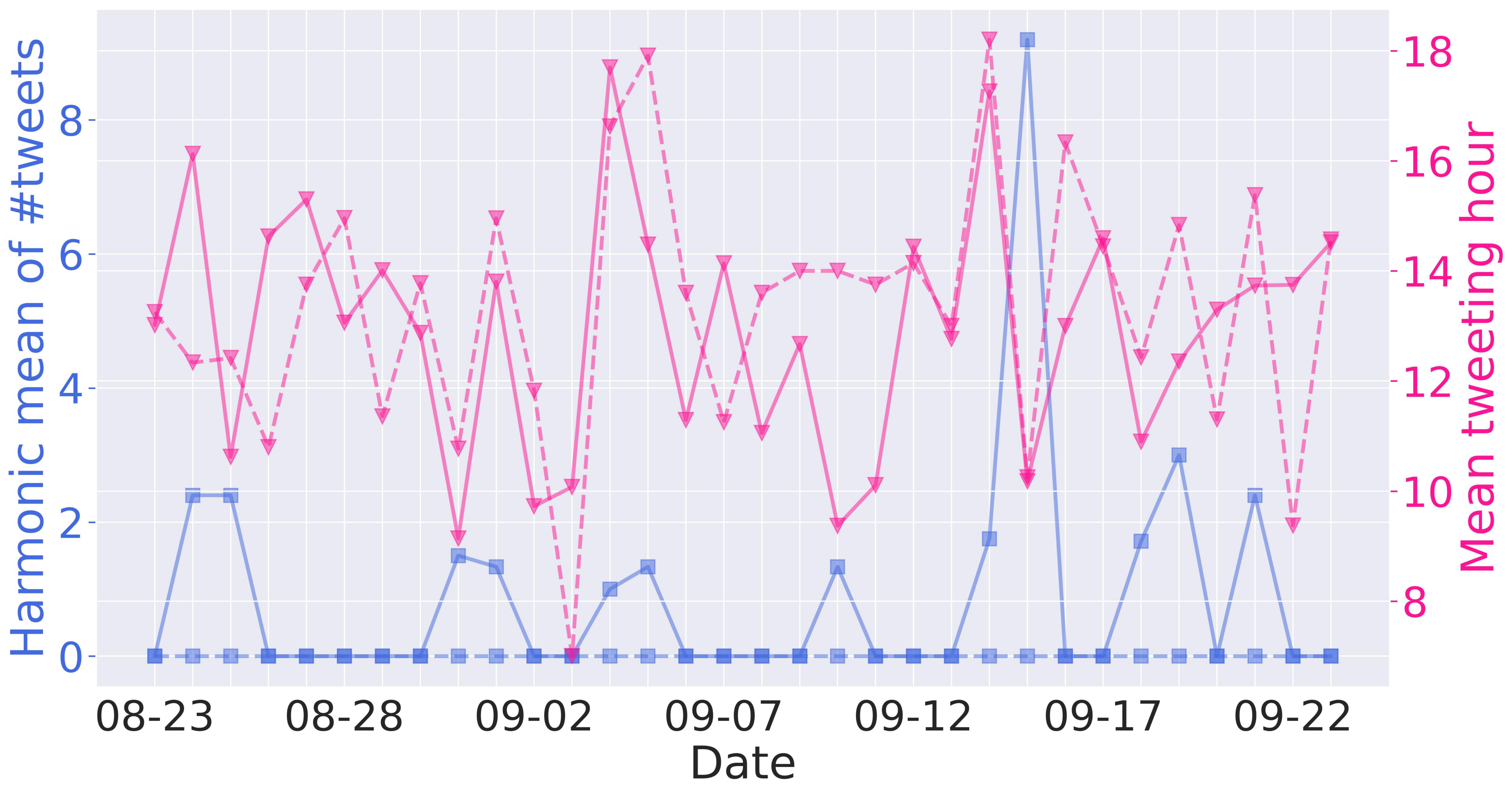}
            }
            \subfigure[Meloni] 
            {
                \label{Meloni_Slope}
                \includegraphics[width=.47\textwidth]{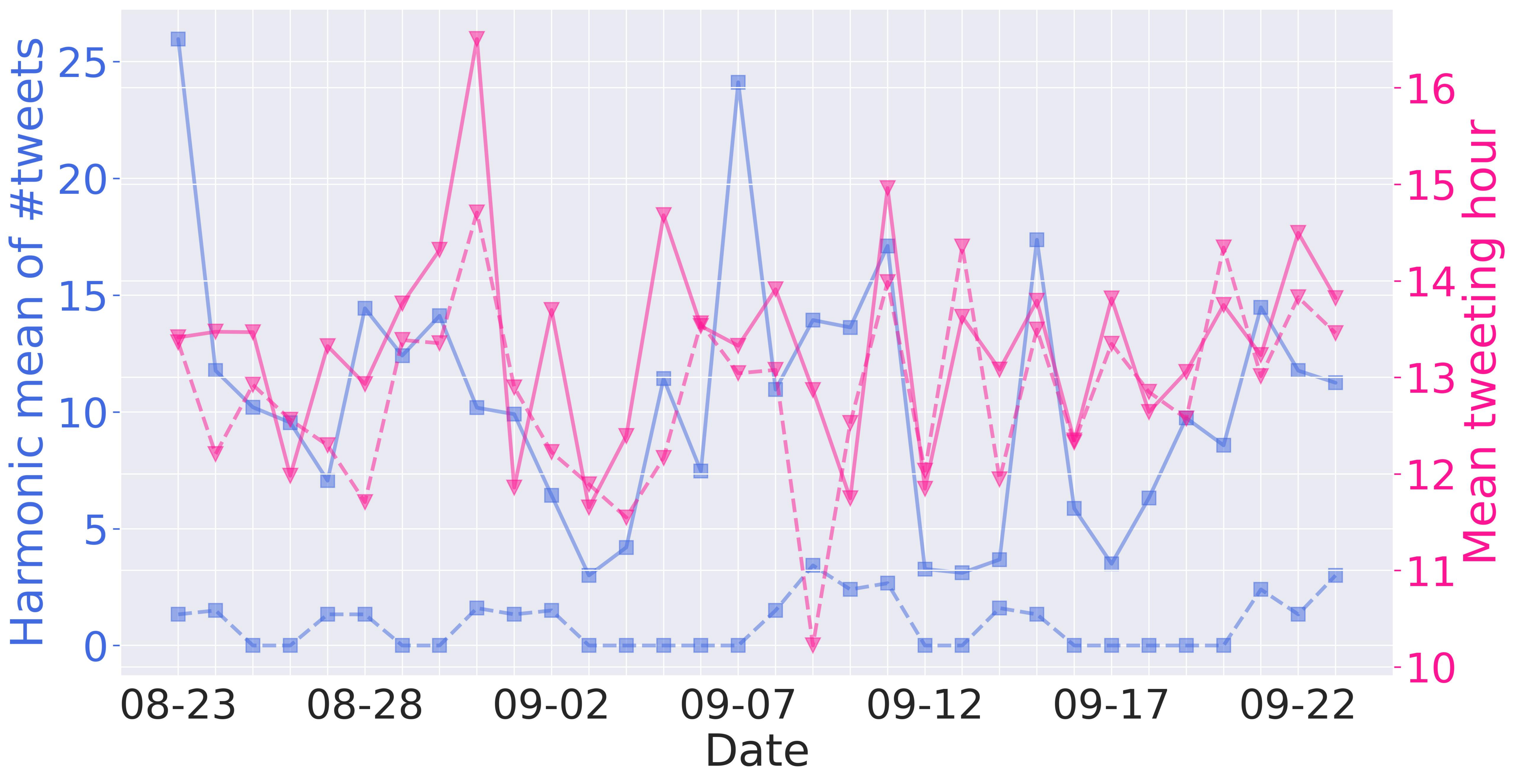}
            }\\
            \subfigure[Salvini] 
            {
                \label{Salvini_Slope}
                \includegraphics[width=.47\textwidth]{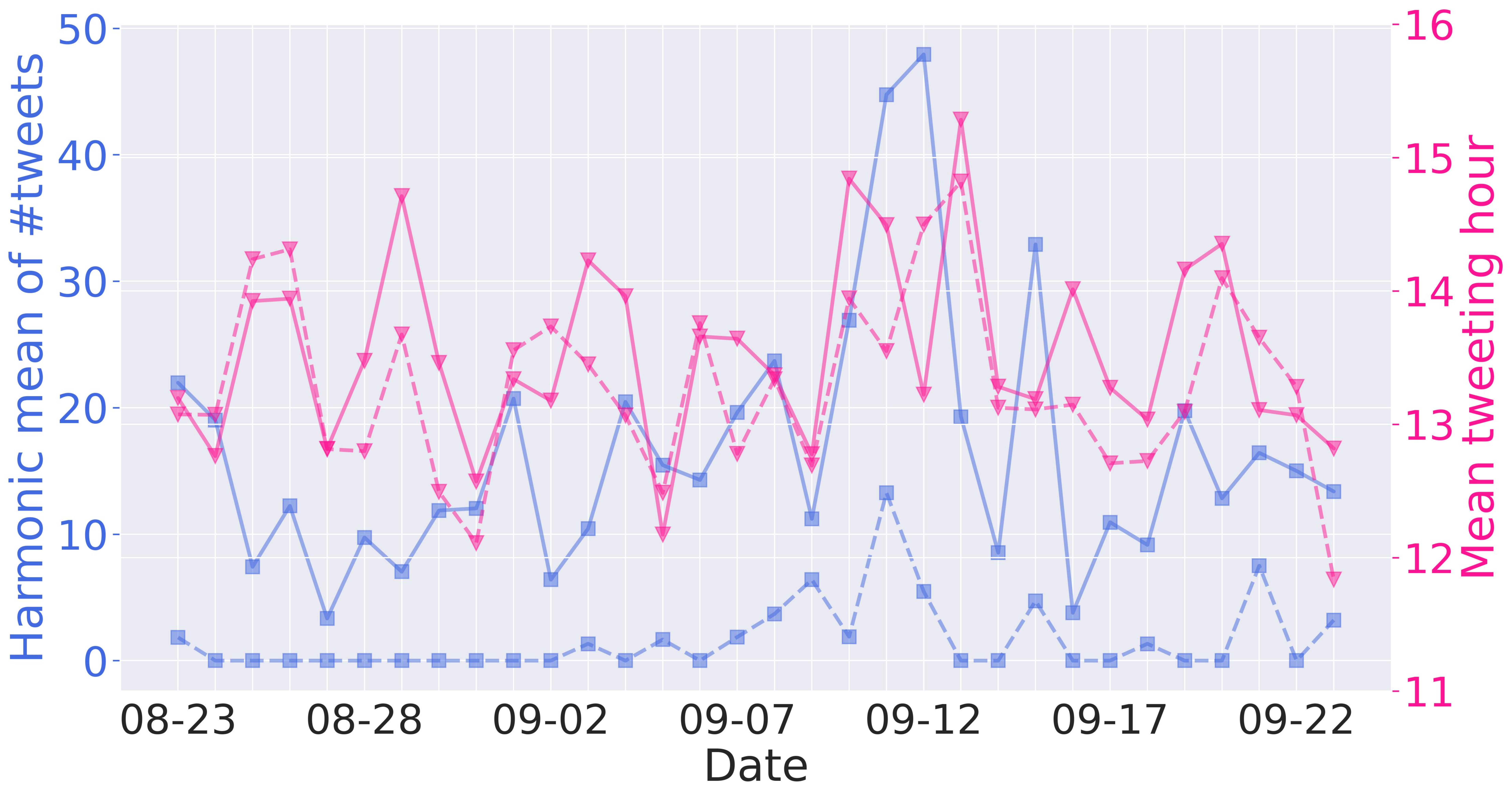} 
            } 
            \subfigure[Berlusconi] 
            {
                \label{Berlusconi_Slope}
                \includegraphics[width=.47\textwidth]{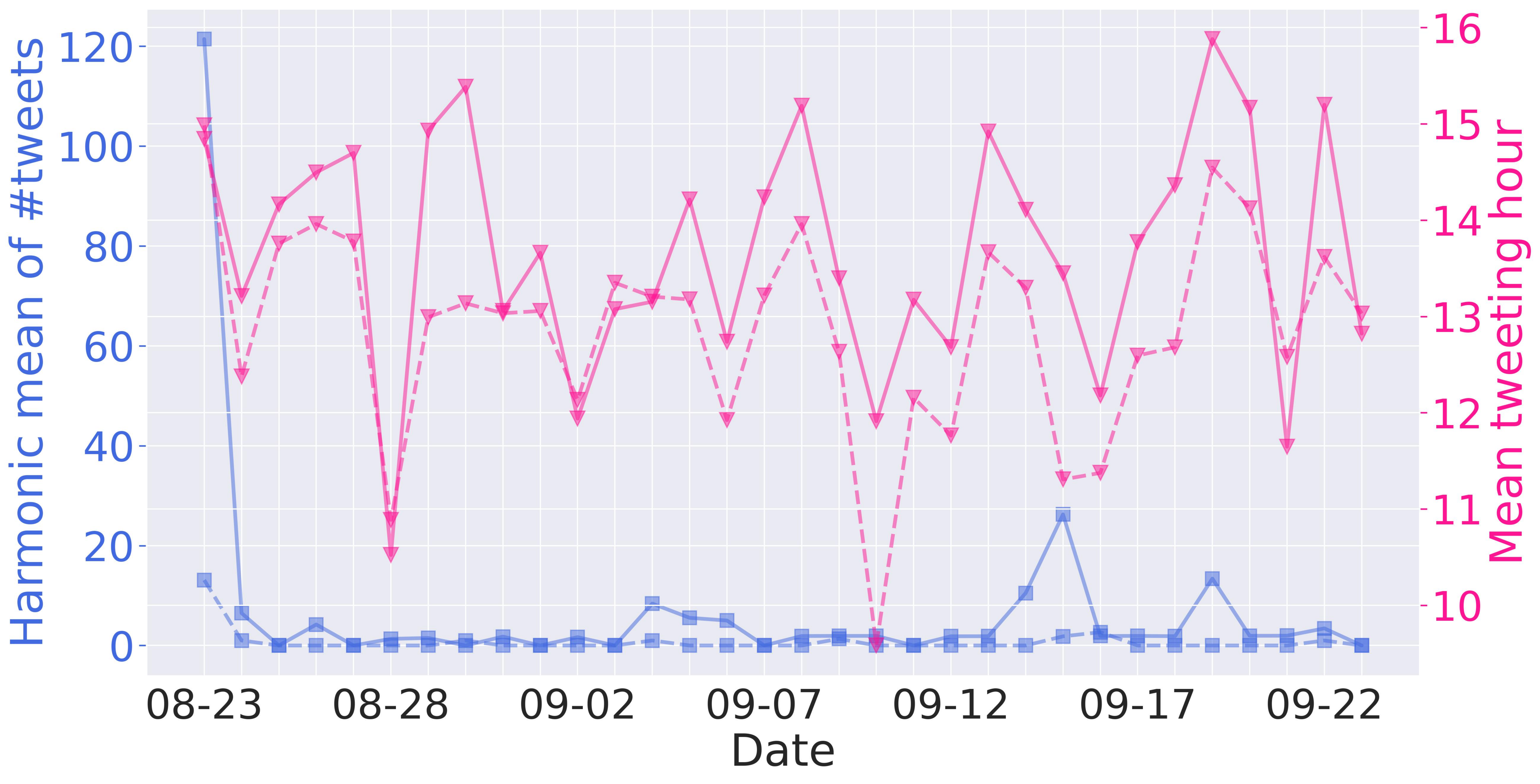} 
            }
            \caption{Mean number of posts and mean posting time for war-related tweets in the last month of Italian elections. Data are reported for both real accounts and bots. }
        \label{fig:SlopeCharts}
\end{figure}

To conclude our analysis, we deeply investigated the final month of the Italian elections, exploring the different discussions and perspectives surrounding the war that emerged under the leaders' posts. 
Our goal is to understand whether humans or bots discussed more the conflict, and which side influenced (or started) the debate. To this aim, we plot a two-scale graph for each party, considering the mean number of tweets concerning the war and the mean posting time (hour) for bots and real users. The results are shown in Figure~\ref{fig:SlopeCharts}.

\vspace{2em}

We computed the \texttt{harmonic\_mean} with the Formula~\ref{eq:harmonic}:
\vspace{0.2cm}
\begin{equation}
    h_{\text{freq}} = 2\times\dfrac{\text{Ukraine\_frequency}\times\text{Russia\_frequency}}{\text{Ukraine\_frequency}+\text{Russia\_frequency}}
    \label{eq:harmonic}
\end{equation}
\vspace{0.2cm}

as an indicator to visualize the number of tweets posted daily by both real users and bots during the last month of the political campaign. Ukraine and Russia posts included only strictly related words to the countries, e.g., ``Ukraine'', ``Ukrainian'', ``Zelensky'', and, ``Russia'', ``Russian'', ``Putin''. 

This measure is bounded from above by the arithmetic mean, indicating its tendency to mitigate the influence of large outliers while accentuating the effect of small ones. This property allows for the evaluation of even the smallest frequencies to be computed, which may be otherwise masked by the influence of dominant outliers in the data. In this scenario, e.g., the results for a politician like Fratoianni, which has a smaller frequency of bots if compared to the other figures, are not suppressed, but his mean will clamp to 0. The other indicator we considered is the \texttt{mean\_tweeting\_hour}, which gives us the arithmetical average of posting time by both genuine and bot accounts.

We focus our attention on the blue spikes in the graphs, which indicate a quantitative increment in the number of tweets regarding the war. The majority of the spikes, either regarding the real or the fake users, concentrate on the period between 10 September and 24 September. The number of tweets posted by real users is always greater than bots' posts, which is in accordance with the percentage of bots found earlier ($\sim$12\%). Looking at the \texttt{mean\_tweeting\_hour},  we can establish that on various occasions the bots posted tweets in a time before the spikes coming from the real users, on average. This trend is glaring for Conte, Meloni, Salvini, and Berlusconi, in which bots often started tweeting before the real users, hence influencing or driving the daily discussion.



\section{Discussion}
\label{sec:discussion}

Our analyses found that Italian politics has actively considered the Russo-Ukrainian conflict in their campaigns, with parties taking on a greater role than others. Additionally, we found a fair number of bots to be active and influential during the last elections. The effect seems to be tied to the particular parties or coalitions, requiring further investigation. Indeed, we could not determine nor speculate on who was driving these bots or for what purpose. 
Anyhow, our findings demonstrate that external events can significantly impact local (national) ones, with unpredictable consequences. Social media platforms like Twitter are credited with democratizing discussions about politics and social issues, but as demonstrated in the literature, manipulation of information is an actual threat rather than a risk. Unfortunately, most studies addressing this issue focus on English-based data or countries, since state-of-the-art models are more reliable. However, analyzing non-English countries is of utmost interest nowadays, since every country has a significant impact on global political equilibrium. 

As we found interferences in the political scenarios, bots or fake accounts might likely be involved in disinformation or other malicious activities in the country. With the rapid development of Artificial Intelligence, it could always become harder to detect these colluding entities. It is, therefore, necessary to conduct further studies to address the language-specific obstacles, as well as to identify who operates such bots to eventually detect their objectives and contrast them. 

\subsection{Limitations}
As we mentioned earlier, our study was limited by the few models available to process the Italian language. However, we think our work can stimulate further research and improve NLP models for Italian, as well as other minor languages.
An additional limitation relies on the use of the external tool Botometer for the detection of bots. As such, the reliability of our findings is contingent on the accuracy of this tool~\cite{rauchfleisch2020false}. However, Botometer is widely recognized as a state-of-the-art bots detection mechanism, and we have taken a conservative approach in the detection phase to limit false positives. Indeed, the number of bots and their influence could be higher than our estimates, stressing the need for more research in the area.



\section{Conclusion and Future Works}
\label{sec:concl}
The purpose of this study was to investigate how Italian politics responded to the Russian-Ukrainian conflict on Twitter and understand the bots' influence and manipulations before the 2022 general elections in Italy.  Our findings suggest that bots are a significant presence in political conversations on Twitter, with approximately $12\%$ of commenters being identified as bots. We also analyzed the timing in which the bots posted concerning when the real users posted, and we can infer that in some cases, these accounts could have forced a certain direction in the topics discussed online. This highlights the potential impact of automated accounts on public opinion during political campaigns.

Our analysis can be improved in the future in several ways. For instance, we could consider the presence of comments in other languages. As our study focused solely on comments posted in the Italian language, taking into account comments in other idioms could offer a more comprehensive understanding of the discussion. In addition, users' attitudes and behaviors could be studied based on their location, in order to analyze potential regional differences in the discussion. Notably, identifying the geographical location of bots can provide more insight into \textit{who} attempts to manipulate discussion and \textit{why}.

\bibliography{main.bib}{}
\bibliographystyle{splncs04}
\end{document}